\documentclass[11pt,letterpaper,usenames,dvipsnames]{article}
\usepackage{jheppub}

\allowdisplaybreaks[2]  
\usepackage{pifont}
\newcommand{\cmark}{\text{\ding{51}}}
\newcommand{\xmark}{\text{\ding{55}}}
\usepackage{bm} 
\usepackage{dsfont} 
\usepackage{ytableau}   

\usepackage{tikz} 
\usetikzlibrary{arrows,snakes,shapes.arrows,decorations.markings}
     \tikzset{>=triangle 90}
     \tikzstyle{gr}=[draw,circle,green!50!black,fill=green!50!black,scale=.6]
     \tikzstyle{Bl}=[draw,circle,blue,scale=.6]
     \tikzstyle{R}=[draw,circle,fill=red,scale=.6]
     \tikzstyle{bl}=[draw,circle,fill=black,scale=.35]
     \tikzstyle{bbc}=[draw,circle,fill=black,scale=.75]
     \tikzstyle{bbcs}=[draw,circle,fill=black,scale=.5]
     \tikzstyle{rc}=[circle,fill=red,scale=.6]
     \tikzstyle{wc}=[draw,circle,scale=.75]
     \tikzset{
    clip even odd rule/.code={\pgfseteorule}, 
    invclip/.style={
        clip,insert path=
            [clip even odd rule]{
                [reset cm](-\maxdimen,-\maxdimen)rectangle(\maxdimen,\maxdimen)
            }
        }
    }

\usepackage{array} 
\usepackage{multirow} 
\usepackage{colortbl} 
\usepackage{arydshln} 
\usepackage{rotating} 

\usepackage{xcolor}   
\def\blue#1{{\color{blue}{#1}}}
\def\green#1{{\color{black!25!green}{#1}}}

\def\rcy{\rowcolor{black!25!yellow!10}}

\def\rcr{\rowcolor{red!07}}

\usepackage{url} 

\newcommand{\beq}{\begin{equation}}
\newcommand{\eeq}{\end{equation}}
\def\del{{\partial}}
\def\bar{\overline}
\def\til{\widetilde}
\def\hat{\widehat}
 
\def\vev#1{{\langle{#1}\rangle}} 

\def\^{\wedge}

\def\U{{\rm U}}
\def\SU{{\rm SU}}

\def\SO{\mathop{\rm so}}

\def\Sp{\mathop{\rm sp}}

\def\with{\mathop{\,\textstyle{\rm w/}\,}}
\def\yd{\ydiagram}
\def\C{\mathbb{C}} 
\def\H{\mathbb{H}}
 
\def\P{\mathbb{P}} 
\def\R{\mathbb{R}} 
\def\Z{\mathbb{Z}} 
\def\ff{{\mathfrak f}}
\def\gf{{\mathfrak g}}

\def\uf{{\mathfrak u}}
\def\bd{{\bf d}}

\def\tj{{\til\jmath}}

\def\bm{{\bf m}}

\def\bn{{\bf n}}

\def\br{{\bf r}}

\def\bv{{\bf v}}

\def\cA{{\mathcal A}}
\def\cB{{\mathcal B}}
\def\cBh{{\hat\cB}}
\def\cC{{\mathcal C}}

\def\cD{{\mathcal D}}
\def\cE{{\mathcal E}}
\def\cH{{\mathcal H}}
\def\cI{{\mathcal I}}
\def\cJ{{\mathcal J}}

\def\cN{{\mathcal N}}
\def\cO{{\mathcal O}}

\def\cS{{\mathcal S}}

\def\cW{{\mathcal W}}

\def\a{{\alpha}}

\def\b{{\beta}}

\def\g{{\gamma}}

\def\D{{\Delta}}
\def\e{{\epsilon}}

\def\th{{\theta}}

\def\l{{\lambda}}

\def\x{{\xi}}

\def\s{{\sigma}}
\def\S{{\Sigma}}

\def\ch{{\chi}}
\def\f{{\phi}}
\def\vf{{\varphi}}

\title{Geometric constraints on the space of N=2 SCFTs\\ 
III: enhanced Coulomb branches and central charges}
\author{Philip Argyres,}
\author{Matteo Lotito,}
\author{Yongchao L\"u,}
\author{and Mario Martone}
\affiliation{Physics Department, University of Cincinnati,\\
Cincinnati OH 45221-0011, USA}
\emailAdd{philip.argyres@gmail.com}
\emailAdd{lotitomo@mail.uc.edu}
\emailAdd{lychaoaa@gmail.com}
\emailAdd{martonmo@ucmail.uc.edu}
\abstract{This is the third in a series of three papers on the systematic analysis of rank 1 four dimensional $\cN{=}2$ SCFTs.  In the first two papers \cite{paper1,paper2} we developed and carried out a strategy for classifying and constructing physical planar rank-1 Coulomb branch geometries of $\cN{=}2$ SCFTs.  Here we describe general features of the Higgs and mixed branch geometries of the moduli space of these SCFTs, and use this, along with their Coulomb branch geometry, to compute their conformal and flavor central charges.  We conclude with a summary of the  state of the art for rank-1 $\cN{=}2$ SCFTs.}

\begin{document}
\maketitle


\section{Introduction}

This is the third in a series of three papers on the classification of 4d $\cN=2$ superconformal field theories (SCFTs) and their relevant deformations with rank-1 Coulomb branches with planar topology.  In two previous papers \cite{paper1,paper2} we constructed all consistent planar rank-1 Seiberg-Witten geometries whose generic relevant deformation ``ends" in undeformable singularities\footnote{The completeness of our construction relies on the assumption that no non-trivial rank-0 SCFTs exists. For more details read the conclusion of this paper or the introduction of  \cite{paper1}.}.  Most of these geometries are associated to SCFTs known to exist by other constructions \cite{allm1602, am1604}, but for some the associated SCFT is still conjectural.  We review the status of the known and conjectured rank-1 SCFTs in the conclusion to this paper.

The moduli space of vacua of $\cN=2$ supersymmetric field theories in 4 dimensions can have various intersecting components or ``branches".  Each branch is a complex manifold with singularities.  The Coulomb branch (CB) is the component where complex scalars in $\cN=2$ vector multiplets get vevs, generically have an unbroken $\U(1)^r$ low energy gauge group where $r$ is the complex dimension, or ``rank", of the CB, and have a special K\"ahler geometry with a $\U(1)_R$ holomorphic isometry.  The Higgs branches (HBs) are components where only the quaternionic scalars in hypermultiplets get vevs, generically have no unbroken low energy gauge group, and have hyperk\"ahler geometry with its associated $\SU(2)_R$ isometry.  

Mixed branches are instead varieties where both vector and hypermultiplet scalars get vevs.  By a nonrenormlization theorem \cite{aps96} such a branch is locally metrically a product of a special K\"ahler with a hyperk\"ahler variety.  With one exception, a mixed branch will intersect the CB and any HBs along singular subvarieties of each (see below).  The exception is an ``enhanced Coulomb branch" (ECB) which is a maximal-dimension mixed branch that contains the CB as a subvariety and thus intersects the CB along the whole CB.  When this occurs, there is, properly speaking, no longer a ``pure" CB in the theory, as it is subsumed in the ECB: generic coulombic vacua have both non-vanishing vector multiplet and non-vanishing hypermultiplet vevs. This is mathematically described as a hyperk\"ahler manifold (describing hypermultiplet vevs neutral under the gauge group) fibered over generic points of the CB. Also ECBs have isometry groups of the form $\SU(2)_R\oplus \U(1)_R\oplus \ff'$, where $\ff\supset\ff'$ with $\ff$ the flavor symmetry algebra.

In this paper we will focus on Higgs and mixed branches and will describe ways to determine their properties, and also their conformal and flavor central charges. The latter data is related to measures of the number of degrees of freedom of the SCFT and to how they are charged under its flavor symmetry.  When an ECB exists, its properties are important ingredients in the calculation of the conformal and flavor central charges \cite{st08}.  Their contributions to central charges can be computed from the twisted partition function on the ECB, and will be explained in more detail below.

The main way we can determine the properties of the Higgs branch and the ECB of a rank-1 SCFT is from its connection to gauge theories through RG flows or S-dualities \cite{Argyres:1995xn,as07,aw07,aw10} or to class $\cS$ theories \cite{Gaiotto:2009we,Chacaltana:2014nya,Chacaltana:2016shw}.  For instance the Hall-Littlewood index, a particular limit of the super-conformal index of four-dimensional $\cN=2$ SCFTs \cite{Beem:2014rza}, counts Higgs branch operators and allows the determination of the Higgs data.    Alternatively in the cases of the RG flows between rank-1 SCFTs, described in \cite{allm1602, am1604}, a simple ansatz leads to a consistent description of the chiral ring of the ECB.  In these cases we can also determine the dimension of the Higgs branch and of the ECB, and the flavor symmetry action on the ECB.  With this data, the conformal and flavor algebra central charges can be computed following \cite{st08}.  A summary of our results for a subset of the possible rank-1 SCFTs is shown in table \ref{tab1}.

\begin{table}[ht]
\centering \small
$\def\arraystretch{1.0}
\begin{array}{clcc|c|cc:cc|ccc:cc}
&\multicolumn{3}{l|}{\text{CB:}} &
\multicolumn{1}{l|}{\text{HB:}} &
\multicolumn{3}{l}{\text{ECB \&\ flavor symm.:}} & \Z_2 &
\multicolumn{5}{l}{\text{Central charges:}} 
\\[1mm]
&\text{SI sing.} & \D(u) &\text{deform.} 
&\ \ d_{\text{HB}}\ \  
&\ \  h\ \ &\ \ {\bf 2h}\ \ &\quad \ff\quad &\text{obst?}
&\ \ k_\ff\ \ &\ \ 24a\ \ &\ \ 12c\ \ 
&\ \ b\ \ &\ \ e\ \ 
\\[1.5mm]
\hline\hline
&&&&&&&&&&&\\[-4.5mm]
&II^* & 6 & \{{I_1}^{10}\}
& 29 & 0 & - & E_8 & -
& 12 & 95 & 62 & 20 & 1\\
&III^* & 4 & \{{I_1}^9\}
& 17 & 0 & - & E_7 & \xmark
& 8 & 59 & 38 & 18 & 1\\
&IV^* & 3 & \{{I_1}^8\}
& 11 & 0 & - & E_6 & -
& 6 & 41 & 26 & 16 & 1\\
\rcy &I_0^* & 2 & \{{I_1}^6\} 
& 5 & 0 & - & D_4 & -
& 4 & 23 & 14 & 12 & 1\\
&IV & 3/2 & \{{I_1}^4\} 
& 2 & 0 & - & A_2 & -
& 3 & 14 & 8 & 8 & 1\\
&III & 4/3 & \{{I_1}^3\} 
& 1 & 0 & - & A_1 & \xmark
& 8/3 & 11 & 6 & 6 & 2\\
&II & 6/5 & \{{I_1}^3\} 
& 0 & 0 & - & \varnothing & -
& - & 43/5 & 22/5 & 4 & - \\
\rcr \multirow{-9}{4mm}{\begin{sideways}$I_1$ series\qquad \  \end{sideways}}
&I_1 & 1 & - 
& 0 & 0 & - & U_1 & -
& * & 6 & 3 & 2 & - \\[.5mm]
\hline\hline
&&&&&&&&&&&\\[-4.5mm]
&II^* & 6 & \{{I_1}^6,I_4\} 
& 16 & 5 & \bf10 & C_5 & \cmark
& 7 & 82 & 49 & 14 & \frac12\\
&III^* & 4 & \{{I_1}^5,I_4\} 
& 8 & 3 & (\bf6,1) & C_3A_1 & (\cmark,\xmark)
& (5,8) & 50 & 29 & 12 & (\frac12,1)\\
&IV^* & 3 & \{{I_1}^4,I_4\} 
& 4 & 2 & {\bf4}_0 & C_2U_1 & (\cmark,-)
& (4,?) & 34 & 19 & 10 & (\frac12,-)\\
\rcy &\blue{I_0^*} & 2 & \{{I_1}^2,I_4\} 
& 0 & 1 & \bf 2 & C_1 & \cmark
& 3 & 18 & 9 & 6 & \frac12\\
\rcr \multirow{-6}{4mm}{\begin{sideways}$I_4$ series\qquad \ \end{sideways}}
&I_4 & 1 & - 
& 0 & 0 & - & U_1 & - 
& * & 6 & 3 & 2 & - \\[.5mm]
\hline\hline
&&&&&&&&&&&\\[-4.5mm]
&II^* & 6 & \{{I_1}^3,I^*_1\} 
& 9 & 4 & \bf 4 \oplus \bar 4 & A_3{\rtimes}\Z_2 & -
& 14 & 75 & 42 & 12 & 1\\
&III^* & 4 & \{{I_1}^2,I^*_1\} 
& ? & 2 & \bf 2_+ {\oplus}\, 2_- & A_1U_1{\rtimes}\Z_2 & (\xmark,-)
& (10,?) & 45 & 24 & 10 & (1,-)\\
&\green{IV^*} & 3 & \{I_1,I^*_1\} 
& 0 & 1 & \bf 1_+ {\oplus}\, 1_- & U_1 & -
& * & 30 & 15 & 8 & -\\
\rcr \multirow{-5}{4mm}{\begin{sideways}$I_1^*$ series\qquad\  \end{sideways}} 
&I_1^* & 2 & - 
& 0 & 0 & - & \varnothing & -
& - & 17 & 8 & 6 & -\\[.5mm]
\hline\hline
&&&&&&&&&&&\\[-4.5mm]
&II^* & 6 & \{{I_1}^2,IV^*_{Q=1}\} 
& ? & 3 & \bf 3 \oplus \bar 3 & A_2{\rtimes}\Z_2 & -
& 14 & 71 & 38 & 11 & 1\\
&\green{III^*} & 4 & \{I_1,IV^*_{Q=1}\} 
& 0 & 1 & \bf 1_+ {\oplus}\, 1_- & U_1{\rtimes}\Z_2 & -
& * & 42 & 21 & 9 & -\\
\multirow{-4}{4mm}{\begin{sideways}$\scriptstyle{IV^*_{\scriptscriptstyle Q=1}}$ \small{ser.}\quad \  \end{sideways}} 
&IV^*_{Q=1} & 3 & - 
& 0 & 0 & - & \varnothing & -
& - & 55/2 & 25/2 & 7 & -\\[.5mm]
\hline\hline
\rcy &&&&&&&&&&&\\[-4.5mm]
\rcy &\blue{I_0^*} & 2 & \{{I_2}^3\} 
& 0 & 1 & \bf 2 & C_1 & \cmark
& 3 & 18 & 9 & 6 & \frac12\\
\rcr \multirow{-3}{4mm}{\begin{sideways}$I_2$ ser.\quad\ \  \end{sideways}} 
&I_2 & 1 & - 
& 0 & 0 & - & U_1 & -
& * & 6 & 3 & 2 & -\\[.5mm]
\end{array}$
\caption{Partial list of rank-1 $\cN=2$ SCFTs.  They are divided into 5 series; the CFTs within each series are connected by RG flows from top to bottom.  The red rows give the characteristic IR-free theory each series flows to.  Yellow rows are lagrangian CFTs, while blue and green singularities have enhanced $\cN=4$ and $\cN=3$ supersymmetry, respectively.  The first 3 columns describe the CB geometry; the next column gives the HB dimension; the next 4 columns give properties of the ECB and the flavor symmetry; and the last five columns give the CFT central charges.  The meaning of each column and the choice of the theories appearing in the rows are explained in the introduction.\label{tab1}}
\end{table}

Table \ref{tab1} does not list all the possible planar rank-1 CB geometries but only those for which there is independent evidence for the existence of an associated SCFT.  There are, for instance, non-listed geometries associated to gauging discrete symmetries of many of the theories in table \ref{tab1}, and which give rise to distinct CB geometries, described in \cite{am1604}.  The central charges of a theory and of any of its discretely gauged versions are the same, though the flavor symmetry and HB and ECB fibers may change, again as described in \cite{am1604}. Table \ref{tab1} also includes some IR-free theories that the SCFTs flow into upon turning on relevant deformations (i.e., mass terms or chiral deformation terms).  

The SCFTs in the table are arranged into 5 series: the theories within each series are related by RG flows from the topmost line to the bottom.  These flows are described in more detail in \cite{paper2,allm1602}.   The first 3 columns of table \ref{tab1} give the Kodaira type of the scale-invariant singularity on the CB, the scaling dimension of the local coordinate on the CB, and the deformation pattern of the singularity under deformation by generic relevant operators, respectively.  This data is discussed in great detail in \cite{paper1, paper2}.   The $d_\text{HB}$ column gives the quaternionic dimension of the Higgs branch where it can be determined, while the next four columns give properties of the ECB and flavor symmetry: $h$ is the quaternionic dimension of the ECB fiber, $\ff$ is the flavor symmetry,\footnote{We use Dynkin notation for simple Lie algebras together with ``$U_1$" to denote $\U(1)$ factors
.} $\bf 2h$ is the representation of $\ff$ under which the ECB fiber transforms, and the last column records whether or not there is a $\Z_2$ global anomaly obstruction \cite{w82} to gauging those simple flavor factors which have symplectic representations.  The last five columns of the table record the flavor, $k_\ff$, and conformal, $a$ and $c$, central charges; $b$ and $e$ are combinations of $k_\ff$, $a$ and $c$ defined in section \ref{sec3} which must satisfy certain integrality constraints.  We do not determine the flavor central charges of $\U(1)$ flavor factors: those marked with a question mark because we do not have enough information to do so, while those marked with a star can be determined but whose values only have meaning relative to some arbitrary conventional normalization (discussed in section \ref{sec3.3}).

The SCFT interpretation of some geometries in table \ref{tab1}, as well as some others not shown in the table, is not unique, as described in \cite{paper2,allm1602}.  These alternative SCFTs depend on intepretations of various undeformable $I_n^*$, $IV^*$, and $III^*$ singularities \cite{paper1} which occur at the end of RG flows as certain ``frozen" interacting SCFTs.  The techniques of this paper can be used to put constraints on the central charges and HB and ECB dimensions of these (hypothetical) theories, but are not powerful enough to determine them completely.


The paper is organized as follows.  
Section \ref{sec2} describes some general properties of Higgs branches of $\cN=2$ field theories, and focuses, in particular, on those of rank-1 SCFTs.  Appendix \ref{sec2.1.2} shows that for gauge theories only hypermultiplets in representations of the gauge group in the same center conjugacy class as the adjoint representation can give rise to ECBs.  These are always orthogonal representations (though the converse does not hold) so have a flavor symmetry, $\ff\supset\ff'$, which contains a semisimple symplectic factor, $\ff'$, which acts faithfully on the ECB.  Strongly-coupled superconformal theories for which there is no known lagrangian description can also have ECBs.  This follows in many cases from S-duality arguments, as described in section \ref{sec2.2} for the particular case of rank-1 SCFTs.  For these theories we determine the quaternionic dimension of the HB, $d_{\text{HB}}$, and that of the ECB fiber, $h$, as well as the action of the flavor symmetry on the ECB.  In section \ref{sec2.3} we determine the chiral ring of the ECBs for these theories.

Since the local properties of an ECB follow from $\cN=2$ supersymmetric nonrenormalization theorems, it follows with only mild assumptions that the argument of \cite{st08} relating CFT data (CB scaling dimensions, flavor symmetry, central charges) to the geometry of the CB near a singularity can be extended to the case with an ECB, which we do in section \ref{sec3}.  In \cite{paper1, paper2} we explained how under relevant deformations a singularity associated with a given SCFT splits into lesser ones, also interpreted as SCFTs.  The basic idea then is to determine the topologically twisted partition function on the ECB of the initial SCFT by relating it to the partition functions of these lesser SCFTs.  For the case of rank 1 CBs this gives definite relations, and allows us to compute the $a$ and $c$ conformal central charges, and the current algebra central charges, $k$, for semisimple factors of the flavor symmetry.   Finally, in section \ref{sec3.4}, we discuss constraints on rank-1 SCFTs coming from the various central charge inequalities appearing in the literature, and new constraints coming from integrality conditions on central charges arising from single-valuedness of the ECB partition function measure.  While the former does not give strong constraints, the latter has an interesting relationship to the existence of $\Z_2$ obstructions to gauging flavor symmetries and to the global form of the flavor symmetry group.

In the concluding section \ref{sec4}, we summarize the results of \cite{paper1,paper2,allm1602} and this paper presenting an encompassing picture of the status of the art for $\cN=2$ rank-1 SCFTs.   We then discuss the evidence for which of the constructed rank-1 CB geometries correspond to actual SCFTs, and formulate a conservative conjecture that the only rank-1 $\cN=2$ SCFTs with planar CBs are those shown in table \ref{tab1} together with their discrete gaugings listed in \cite{am1604}.   We end with a list of some open questions.


\section{Moduli space of rank-1 SCFTs}
\label{sec2}

In this section we develop aspects of the complex and metric structure of the moduli space of $\cN=2$ superconformal gauge and non-lagrangian theories.  In the rank-1 case only two branches are possible: an enhanced Coulomb branch (ECB) and a Higgs branch (HB).

In favorable situations there is a way to determine the HB and ECB data of strongly-coupled, isolated, $\cN=2$ SCFT through their S-dual description. This is the case if by weakly gauging part of the flavor symmetry of the isolated rank-1 theory we can construct a higher-rank scale invariant $\cN=2$ SCFT with a SUSY gauge theory dual description \cite{as07}. We will use such S-dual descriptions for most of the SCFTs in table \ref{tab1}.  For the others, this data is deduced from class $\cS$ techniques \cite{Gaiotto:2009we,Beem:2014rza,Rastelli:2014jja,Chacaltana:2016shw}, or from the assumption of $\cN=3$ supersymmetry \cite{ae1512,gr1512,nt1602,at1602}, as described in \cite{allm1602}.
  
Finally, assuming the isomorphism between coordinate rings of $\cN=2$ moduli space and chiral rings of ECB and HB operators in the SCFT \cite{Beem:2014zpa}, we give a description of many of these branches as coadjoint orbits of the flavor algebra.

\subsection{Higgs branches (HBs) and mixed branches}
\label{sec2.1}

We start by reviewing the general structure of the moduli space of vacua of $\cN=2$ field theories that follows from the selection rules of unbroken supersymmetry;  see, e.g., \cite{aps96}.  As already mentioned above, the general branch of the moduli space of an $\cN=2$ field theory is one where there are both $n_v$ massless vector multiplets and $n_h$ massless neutral hypermultiplets.  This branch is locally metrically a cartesian product of an $n_v$-complex-dimensional special K\"ahler manifold with an $n_h$-quaternionic-dimensional hyperk\"ahler manifold.  Furthermore, the hyperk\"ahler metric cannot depend on any masses or chiral deformation parameters (relevant or marginal) of the field theory.\footnote{$\cN=2$ Fayet-Iliopoulos terms can deform hypermultiplet moduli spaces (see, e.g., \cite{Antoniadis:1996ra} for a review), but do not occur as $\cN=2$ supersymmetric deformation parameters of $\cN=2$ SCFTs \cite{paper1};  they can occur in $\cN=2$ supergravity theories, however.}  This implies, in particular, that the hyperk\"ahler factors are scale-invariant, and thus are metrically cones \cite{gr98,wkv99}.

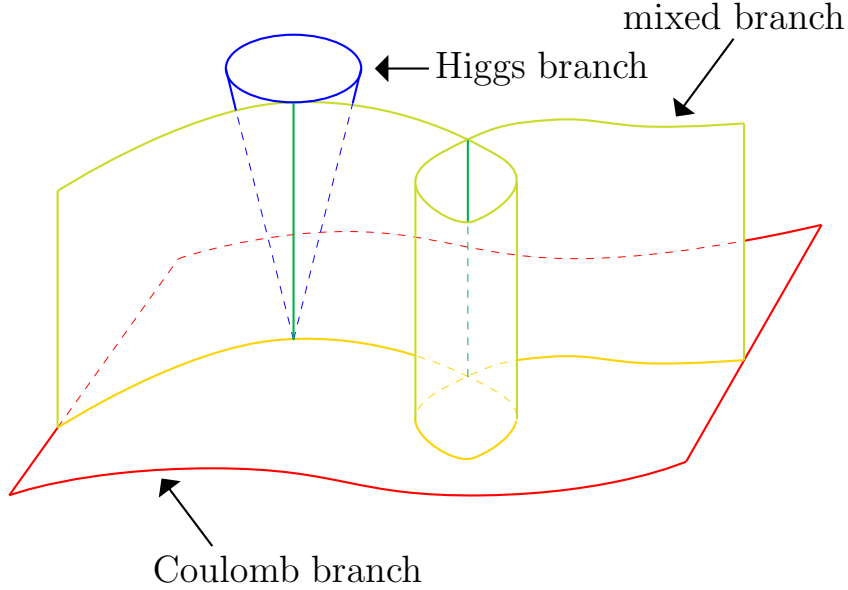
\begin{figure}[htb]
\centering

\begin{tikzpicture}[thick, scale=0.45]

\draw [red] (-12,-3)  -- (-10.57,-1);  
\draw [red, ultra thin, dashed] (-10.57,-1)  -- (-7,4);  
\draw [red] (8,-2) -- (12,5);  
\draw [red] plot [smooth, tension=1] coordinates { (-12,-3) (-6,-2.2) (2,-3)(8,-2)}; 

\begin{scope}
\clip (-10.57,3) rectangle (9.71,5);
\draw [red, ultra thin, dashed] plot [smooth, tension=0.8] coordinates { (-7,4) (-2,4.8) (5,4)(12,5)};   
\end{scope}

\begin{scope} 
\begin{pgfinterruptboundingbox}
\path[invclip] (-10.57,3) rectangle (9.71,5);
\end{pgfinterruptboundingbox}
\draw [red] plot [smooth, tension=0.8] coordinates { (-7,4) (-2,4.8) (5,4)(12,5)};   
\end{scope}

\begin{scope}
\clip (0,-0.8) rectangle (3,2.5);
\draw [yellow!85!red, ultra thin, dashed] plot [smooth, tension=0.8] coordinates { (-10.57,-1) (-4,1.6) (2.5,0)(1.7,-1.9)(0,-0.8)(1.5,0.5)(4,1.1)(7,0.9)(9.71,1)}; 
\end{scope}

\begin{scope} 
\begin{pgfinterruptboundingbox}
\path[invclip] (0,-0.8) rectangle (3,2.5);
\end{pgfinterruptboundingbox}
\draw [yellow!85!red] plot [smooth, tension=0.8] coordinates { (-10.57,-1) (-4,1.6) (2.5,0)(1.7,-1.9)(0,-0.8)(1.5,0.5)(4,1.1)(7,0.9)(9.71,1)}; 
\end{scope}

\draw [yellow!75!green] plot [smooth, tension=0.8] coordinates { (-10.57,6) (-4,8.6) (2.5,7)(1.7,5.1)(0,6.2)(1.5,7.5)(4,8.1)(7,7.9)(9.71,8)}; 
\draw [yellow!75!green] (-10.57,6)  -- (-10.57,-1);  
\draw [yellow!75!green] (9.71,8) -- (9.71,1);  
\draw [yellow!75!green] (0,-0.8) -- (0,6.2);  
\draw [yellow!75!green] (3,-0.75) -- (3,6.25);  

\draw [blue] (-3.6,9.62) ellipse (2 and 1);  
\draw [blue, ultra thin, dashed] (-3.6,1.6) -- (-1.85,8.6); 
\draw [blue] (-1.85,8.6) -- (-1.6,9.62); 
\draw [blue, ultra thin, dashed] (-3.6,1.6) -- (-5.3,8.38); 
\draw [blue] (-5.3,8.38) -- (-5.6,9.62); 

\draw [blue!30!green] (-3.6,1.6) --(-3.6,8.6); 
\draw [blue!30!green, ultra thin, dashed](1.55,0.5)--(1.55,5.1); 
\draw [blue!30!green] (1.55,7.5)--(1.55,5.1); 

\draw[->](0.4,9.62) node [xshift= 1.5cm,yshift= -0.0cm]{\Large Higgs branch} -- (-1.2,9.62);
\draw[->](9.4,10.5) node [xshift= 0.0cm,yshift= 0.3cm]{\Large mixed branch} -- (7.7,8.2);
\draw[->](-6.0,-4.5) node [xshift= 1.0cm,yshift= -0.3cm]{\Large Coulomb branch} -- (-7.5,-2.5);

\end{tikzpicture}
\caption{Visualization of an $\cN=2$ moduli space.  The hyperk\"ahler directions are vertical and the special K\"ahler directions horizontal, with the different types of branches labelled.  Mixed branches are metrically a cartesian product of hyperk\"ahler and special K\"ahler directions except perhaps over complex codimension one subvarieties of their special K\"ahler base. \label{fig1}}
\end{figure}

If both $n_v$ and $n_h$ are non-zero, these are called mixed branches.  A branch whose generic point has only massless vector multiplets ($n_h=0$) is called the Coulomb branch (CB), while a branch whose generic point has no vector multiplets ($n_v=0$) is called a Higgs branch (HB).  A mixed branch with $(n_v, n_h) = (n_v^\text{mixed}, n_h^\text{mixed})$ intersects the CB along an $n_v^\text{mixed}$-complex-dimensional special K\"ahler subvariety.  It can likewise intersect a Higgs branch along an $n_h^\text{mixed}$-quaternionic-dimensional hyperk\"ahler subvariety.  (Also, mixed branches can intersect each other in both special K\"ahler and hyperk\"ahler directions.)   A cartoon illustrating this kind of moduli space is shown in figure \ref{fig1}.

The combination of the action of the dilations and the $\U(1)_R\times\SU(2)_R$ R-symmetry group, provide a $\C^*$ action on the special K\"ahler factor and an $\H^*$ action on the hyperk\"ahler one. It goes as follows. For a conformal theory dilatations act as a homothety on the moduli space.  Choose coordinates on each mixed branch that diagonalize the dilatation action, with complex $u_a$ coordinates on the special K\"ahler factor and pairs of complex $z^i_k$, $i=1,2$, for the hyperk\"ahler factor.\footnote{In general these coordinates may not be algebraically independent, but will satisfy some homogeneous relations.  For instance, in a theory with a rank-2 CB with homogeneous coordinates $\{u_2,u_3\}$ of dimensions 2 and 3, respectively, there could be a mixed branch over a subvariety of the CB defined by $u_2^3=u_3^2$.}  Then the $\U(1)_R$ acts on the vector multiplet factor by phase rotations $u_a \mapsto e^{i \D(u_a) \a} u_a$ where $\D(u_a)$ is the scaling dimension of $u_a$ while the $\SU(2)_R$ acts on the hypermultiplet factor by rotating $(z^1_k, \bar z^2_k)$ as doublets, or, equivalently, by quaternionic ``phase" rotations. Combining all together we get the $\C^*$ and $\H^*$ action mentioned above.

\subsubsection{Enhanced Coulomb branches (ECBs)}\label{ECBs}

There is a special case of mixed branches which deserves a separate discussion.  These are ones where $n_v^\text{mixed}=n_v^\text{CB}$.  In this case the CB is a subvariety of the mixed branch:  equivalently, there are $n_h^\text{mixed}$ massless neutral hypermultiplets at generic points of the CB.  Thus, in this case the CB is effectively enlarged to an $(n_v^\text{CB} + 2n_h^\text{mixed})$-complex-dimensional space. For this reason we will call such mixed branches ``enhanced Coulomb branches" (ECBs).

Note that the ECB is not singular along the CB subvariety.  This follows because the $n_h^\text{mixed}$ massless hypermultiplets of the ECB are neutral with respect to all the CB $\U(1)$ gauge factors (otherwise giving them a vev would Higgs some of the $\U(1)$s, lifting those CB directions, and thus not be an ECB).  Thus the geometry in a neighborhood of a generic point in the ``root" of the ECB (i.e., where it intersects the CB) is a cartesian product of a CB neighborhood with a smooth hyperk\"ahler manfold describing the hypermultiplets vevs.  But, as mentioned above, the hyperk\"ahler manifold is metrically a cone and hence has no intrinsic scales.  But if a cone is smooth at its tip (i.e., where the hypermultiplet vevs branch off the CB), then the curvature tensor and all its derivatives must vanish there as well, since otherwise there would be an intrinsic scale determined by the non-vanishing curvature invariants.   If the metric is analytic in the radial coordinate about the tip, it then follows that it must in fact be flat everywhere.  Thus, the ECB locally has a direct product geometry $U_i \times \H^h$ where $\{U_i\}$ is an open covering of the regular points of the CB, $\H$ is the flat quaternionic line, and $h :=n_h^\text{mixed}$.  

Since there is no singularity at the origin of $\H^h$, there is nothing metrically special to pick out the CB as a subvariety of the ECB.  We will see below, however, that there can be a global twist of the (local) $\text{CB}\times\H^h$ product which fixes the origin of $\H^h$, and thus picks out $\text{CB}\subset\text{ECB}$ as the zero section of the $\H^h \to \text{ECB} \to \text{CB}$ fibration.

There then follow some general constraints on how a global flavor symmetry can act on the ECB fiber, and the associated effect of mass deformations.  The connected component of the $\H^h$ isometry group which fixes its origin\footnote{Since $\H^h$ is flat, the (connected component of the) full isometry group is the euclidean group $SO(4h)\ltimes\R^{4h}$ which includes translations.  We will see below, when we discuss the global structure of the ECB, that the translations can be ignored for SCFTs.} is $SO(4h) \supset \SU(2)_R \times Sp(2h)$, so the $\H^h$ triholomorphic isometry groups (isometries which preserve the hyperk\"ahler structure) are subgroups of $Sp(2h)$.  Call the flavor symmetry group of the theory $F$, with Lie algebra $\ff$.  If $F$ acts faithfully on the ECB, then we must have $\Sp(2h) \supset \ff$.  If $F$ does not act on the ECB, or if the $2h$ complex scalars of the massless hypermultiplets of the ECB have directions which transform as singlets under $\ff$, then those directions will not be lifted upon turning on masses associated to $\ff$.   More generally those complex scalars, $\f_i$, will transform in some $2h$-dimensional, generally reducible, and necessarily symplectic, representation $R$ of $\ff$.  Then masses $m^a$, which transform in the adjoint of $\ff$, couple to the ECB hypermultiplets as $\int d^2\th\, m^a q_i (t^a)_R q_j J^{ij}$, where we are using an $\cN=1$ superfield notation, $(t^a)_R$ are the generators of $\ff$ in the $R$ representation, and $J^{ij}$ is the symplectic form acting on $R$ inherited from its embedding in $\Sp(2h)$.  For generic masses, this lifts all the ECB hypermultiplets except for those components with vanishing $\ff$ weights.  Chiral deformation parameters do not couple to the hypermultiplets, so these deformations, which include marginal deformations, do not lift the ECB hypermultiplet directions.

We can also say a few things about the global structure of the ECB.  Upon following a path around a singularity on the CB (which is generically in complex codimension 1) the ECB fiber will come back to itself up to an isometry, $\s$, which fixes the origin.  In particular, $\s$ does not include any translations of the ECB fiber.  This is because the R- and flavor symmetries are global internal symmetries of the underlying SCFT, and so must act as compact Lie groups on the local fields \cite{Coleman:1967ad,Maldacena:2011jn}.  By taking the limit approaching the singularity, the ECB fiber over the singular subvariety becomes the flat cone $\H^h/\sim_\s$, where $\sim_\s$ is the identification generated by $\s$.  For it to be hyperk\"ahler, we must have $\s\in Sp(2h)$, i.e., it must be a triholomorphic isometry.

The conical fiber over the singular subvariety must also support an $F$-action continuously related to that on the $\H^h$ fibers away from the singularity.  For this action to be well-defined on the conical fiber, we must have that $\s$ commutes with the $F$-action.   Call $F'\subset F$ the part of $F$ which acts faithfully on the ECB fiber, and $\ff'$ its Lie algebra.    Then the ECB fiber $\H^h$ transforms as a non-trivial $2h$-complex-dimensional representation of $\ff'$.  If the representation is irreducible, then $\s\in F'$ and, since it commutes with all elements of $F'$, it must be in the center of $F'$.  For reducible representations, $\s$ need no longer be in $F'$, and there are more possibilities for its action on the ECB fiber.  We will see examples of this below.

\paragraph{ECBs in $\cN=2$ gauge theories} The moduli spaces of $\cN=4$ theories (viewed as $\cN=2$ theories) are familiar examples of ECBs.  In this case the CB is $\C^r/W$ where $r$ is the rank of the gauge group and $W$ is its Weyl group.  The ECB fiber over regular points of the CB are $\H^r = \C^{2r}$, and the total space of the ECB is $\C^{3r}/W$.  Similar, though non-lagrangian, examples are the moduli spaces of $\cN=3$ SCFTs described in \cite{ae1512, gr1512, nt1602, allm1602, at1602}.  

But ECBs commonly occur in strictly $\cN=2$ gauge theories as well.  A careful, yet slightly technical, analysis of the general form of the $\cN=2$ gauge theory lagrangian, reported in appendix \ref{sec2.1.2}, allows us to determine many properties of ECBs that can arise in $\cN=2$ conformal gauge theories.  Here we only summarize our results:

\begin{itemize}

\item[i)] In an $\cN=2$ gauge conformal field theory, ECBs occur whenever there are hypermultiplets in a representation $R$ of the gauge group which has zero weights (e.g., $\SU(2)$ integer spin representations). It can be shown that such representations are necessarily orthogonal, though the converse is not true.

\item[ii)] In $\cN=2$ gauge theories with hypermultiplets transforming in, generally reducible, representations $R$ of the gauge group, the most general flavor symmetry group is a direct sum of unitary, orthogonal and symplectic factors:\footnote{The unitary factors arise from hypermultiplets transforming in conjugate pairs of complex representations of the gauge group, the orthogonal factors from symplectic representations, and the symplectic factors from pairs of orthogonal representations; see, e.g., \cite{mmw13}.}
\begin{align}\label{}
\ff = \left[\oplus_i \U(\ell_i)\right] \oplus 
\left[\oplus_j\SO(m_j)\right] \oplus 
\left[\oplus_k \Sp(2n_k)\right] ;
\end{align}
ECB's can only occur in the theories with symplectic flavor factors.

\item[iii)] The ECB hyperk\"ahler factor transforms as a direct sum of fundamental representations of (some subset of) the flavor symmetry $\ff$ symplectic factors.

\item[iv)] The ECB fiber over a singularity in the CB is a cone $\H^h/\sim_\s$, with $\s$ a triholomorphic isometry of $\H^h$ which fixes the origin and, for lagrangian SCFTs, is always in the $\Z_2$  center of the appropriate symplectic flavor factor.

\end{itemize}

\paragraph{Some examples.}

ECBs thus occur in gauge theories with massless hypermultiplets in orthogonal irreps which are in the same center conjugacy class as the adjoint irrep.  Thus theories for any gauge group with massless adjoint hypermultiplets will always have an ECB.  With one such hypermultiplet, the theory is scale invariant and $\cN=4$ supersymmetric.  The ECB from the $\cN=2$ perspective is just the whole Coulomb branch from the $\cN=4$ perspective.  With $n>1$ adjoint hypermultiplets the theory is an $\cN=2$ IR-free theory.  For a theory with gauge algebra $\gf$, the adjoint has $r = \text{rank}(\gf)$ zero weights, so the ECB fiber is $\H^{rn}$.  The flavor symmetry is $\Sp(2n)$ under which the ECB fiber transforms as $r$ copies of the $\bf 2n$ irrep.  Upon encircling the codimension-1 singularities in the CB where an $\SU(2)$ subalgebra of $\gf$ is restored, the ECB fiber undergoes the monodromy $\H^{rn} \mapsto -\H^{rn}$ (reflection through the origin).  Thus the ECB fiber degenerates to the hyperk\"ahler cone $\H^{rn}/\Z_2$ over these singularities.  (Over intersections of these singularities, the fibers further degenerate to $\H^{rn}/V$ for $V$ appropriate subgroups of the Weyl group of $\gf$.)   There are no larger Higgs branches over the singularities on the CB, so the ECB is the whole moduli space.

There are also examples of asymptotically free or conformal gauge theories with ECBs.  One simple series are $\SO(N)$ gauge theories with $N_f$ massless hypermultiplets in the $\bf N$ irrep for $N$ odd.  In this case the $\bf N$ has a single zero weight, so the ECB fiber is $\H^{N_f}$.  In this case there are larger-dimension Higgs branches over the singularities in the Coulomb branch which contain the degenerate $\H^{N_f}/V$ ECB fiber as subvarieties \cite{Argyres:1996hc}.  Other examples are $\Sp(2N)$ gauge theory with traceless-antisymmetric hypermultiplets, $F_4$ gauge theory with $\bf 26$'s, and $G_2$ gauge theory with $\bf 7$'s.

\subsection{Moduli space for generic, non-lagrangian, rank-1 SCFTs}
\label{sec2.2}

\begin{figure}[htb]
\centering
\begin{tikzpicture} [thick, scale=0.7]
\filldraw [violet!20, outer color=gray!10,inner color=gray!1] (0,7) ellipse (2 and 1);            
\draw[violet!20, fill=gray!35]
     plot [samples=50,domain=-2:2] (\x, {7- sqrt(1-(\x *\x)/4)})
  -- (1.58,5.468) 
  -- (0,6) 
  -- (-1.546, 5.35)
  --cycle;
\draw[lightgray, fill=brown!5]    
    (-6,3.5) --(0,6)
     -- (0,0) --(-6,-2.5) 
     --cycle;
\draw[lightgray, fill=brown!5]
     (0,6) -- (6,4)
     -- (6,-2) --(0,0)  
     --cycle;
\draw [ultra thick,red] (0,0) -- (-6,-2.5) ;  
\draw [ultra thick,red] (0,0) -- (6,-2);
\draw [ultra thick, blue] (0,0) -- (0,6);   
\draw [violet!80, ultra thin,dashed] (0,0) -- (1.58,5.468); 
\draw [violet!80, ultra thin,dashed] (0,0) -- (-1.546, 5.35); 
\draw [ultra thick,green,dashed] (3, -1) -- (3, 5);  
\draw [decorate,decoration={brace,amplitude=10pt,mirror,raise=4pt},yshift=0pt] 
(7,-2) -- (7,4) node [black,midway,xshift= 1.2cm] {$\text{\Large ECB}$};
\draw[->](6,-3) node [xshift= 0.4cm,yshift= -0.2cm]{\Large CB} -- (5,-1.7);
\draw[->](4,5.5) node [xshift= 0.8cm,yshift= 0.4cm]{{\Large $\cH_\text{ECB} \simeq  \H^h$}} -- (3,4.5);
\draw[->](0.5,-1.1) node [yshift= -0.3cm]{\Large conformal vacuum} -- (0,0);
\draw[->](-3,5.5) node [xshift= -0.9cm]{{\Large $\H^h /\sim_{\sigma}$}} -- (0,4.5);
\draw[->](3,8) node [xshift= 0.5cm]{{\Large $\cH_\text{HB}$}} -- (1.8,7.5);
\end{tikzpicture}
\caption{Moduli space of a planar rank-1 N=2 SCFT. \label{fig2}}
\end{figure}
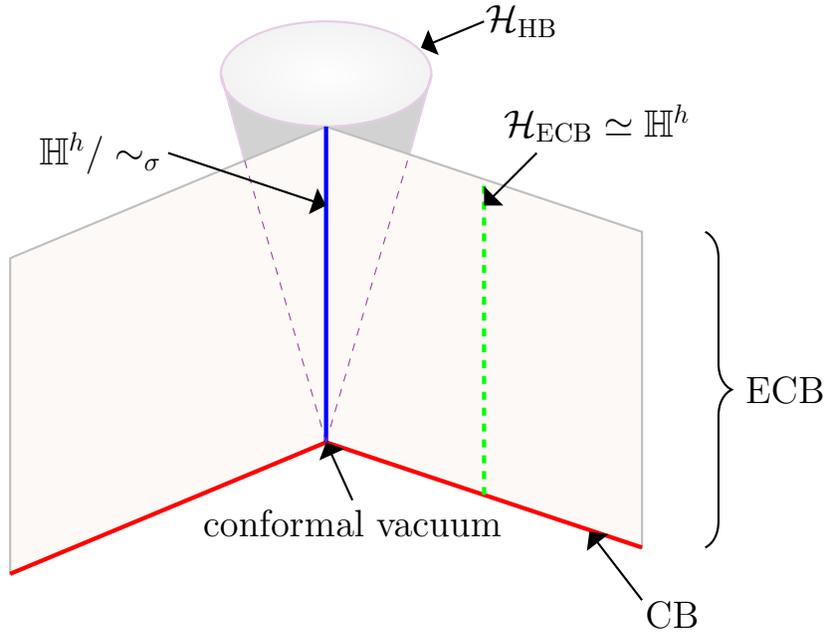

Many of the statements made above are general and apply for any rank. For rank-1 SCFTs, which are the main focus of this paper, the moduli space geometry simplifies considerably and we can use various non-perturbative techniques to extract information about the HB and ECB even for theories which don't have a weak coupling limit. Recall that for planar rank-1 SCFTs the CB geometry is that of a flat 1-complex-dimensional cone with the conformal vacuum at its tip (the ``origin"). The only possible mixed branch is then an ECB with fiber $\cH_\text{ECB}\simeq \H^{h}$ of quaternionic dimension $h$, and there may also be a Higgs branch, $\cH_\text{HB}$, of quaternionic dimension $d_{\text{HB}}$, which is a hyperk\"ahler cone with tip touching the CB at its origin.\footnote{$\cH_\text{HB}$ might have multiple components, and so be a bouquet of cones.}  The intersection of $\cH_\text{HB}$ with the $\cH_\text{ECB}/\sim_\s$ fiber of the ECB over the origin might be any hyperk\"ahler cone from the empty one (the origin istelf) to all of $\cH_\text{ECB}$. In what follows if the only Higgs branch directions over the origin are the $\cH_\text{ECB}/\sim_\s$ fiber, we do not count this as a Higgs branch, and so set $d_{\text{HB}}=0$ in this case.  This general rank-1 moduli space is illustrated in figure \ref{fig2}.

The various techniques which we use to extract information about the moduli spaces of isolated SCFTs are:
\begin{itemize}

\item For those rank-1 SCFTs for which a class $\cS$ construction \cite{Gaiotto:2009we} is available, the Higgs branch can be determined by computing the Hall-Littlewood index \cite{Beem:2014rza}. 

\item We denote an S duality involving gauge theories and rank-1 SCFTs by an equivalence of the form
\begin{align}\label{}
\gf \with {\bf r} = 
\til\gf \with \til{\bf r} \oplus [\text{K},\ff], 
\end{align} 
where the left side stands for an $\cN=2$ vector multiplet with  gauge algebra $\gf$ and massless half-hypermultiplet in gauge representation $\bf r$, and similarly for the right side with a different gauge algebra and hypermultiplet representation, plus a rank-1 SCFT whose CB singularity has Kodaira type ``K" and flavor symmetry $\ff$.  The SCFT is coupled to the $\til\gf$ gauge theory by having a certain $\til\gf\subset\ff$ of the flavor symmetry gauged.

In cases where the SCFT is related to gauge theories by S-duality, the HB and ECB can be determined by extracting this information for the Lagrangian theory on the left side of the duality and asking for consistency with the right side. 

\item For theories which are in the same series in table \ref{tab1}, we can extract HB and ECB data if we know such data for any other theory in the given series. This can be done by carefully following the RG flows which connect them \cite{allm1602}.

\end{itemize}

\noindent Let's now systematically analyze the theories reported in table \ref{tab1}.  Here we will only report a summary of how the entries in \ref{tab1} were computed, we will refer the reader to table \ref{tab1} for the actual numerical values.

\paragraph{\emph{I}$^*_{\bf 1}$ series.} The $II^*$ theory in the ``$I_1^*$ series" in table \ref{tab1}, with flavor symmetry $\ff=\SU(4)\rtimes\Z_2$, is an example of a theory with a class $\cS$ construction which was presented in \cite{Chacaltana:2016shw}.  The ECB for this theory was determined in \cite{allm1602} by matching its central charges.  (The computation of the central charges will be discussed in detail in section \ref{sec3}.)  The ECBs for the other theories in the $I_1^*$ series were then determined by flowing to them from the $II^*$ theory upon turning on suitable masses.

\paragraph{\emph{IV}$^*_{\bf Q=1}$ series.} No class $\cS$ construction is known for the $II^*$ theory with flavor symmetry $\ff=\SU(3)\rtimes\Z_2$ in the $IV^*_{Q=1}$ series.   In this case the ECB was determined in \cite{allm1602} by consistency of its central charges under RG flows under the \emph{assumption} that the $III^*$ theory it flowed to was an $\cN=3$ SCFT. 

Notice that the $I^*_1$ and $IV^*_{Q=1}$ series provide examples of ECBs which are acted upon by unitary, thus non-symplectic, factors of the flavor symmetry. While this cannot occur in gauge theories, as we have seen above, it is allowed in the cases just discussed. In fact the theories in the $I^*_1$ and $IV^*_{Q=1}$ series have no weak coupling limit.

\bigskip

\noindent Both the $I_1$ series and the $I_4$ series (below) have class $\cS$ realizations so their Higgs branch structures can in principle be determined from their Hall-Littlewood index, or, with more work, from the S-duality \cite{Gaiotto:2008nz}.  Here we will focus on determining just the dimension of the Higgs branch and the ECB, which are easy to extract from S-dualities.

\paragraph{\emph{I}$_{\bf 1}$ series.}  The $I_1$ (also known as the maximal deformation) series of SCFTs shown in table \ref{tab1} has been thoroughly studied over the past 20 years, so we simply report their Higgs branch dimensions.  These dimensions are easily computed in the same way as in the $I_4$ series, below; but their Higgs branch structures (e.g., chiral rings) are known explicitly, and coincide with centered 1-instanton moduli spaces.  All the SCFTs in the $I_1$ series have no ECB fiber.

\paragraph{\emph{I}$_{\bf 4}$ series.} In this case we can extract a lot of information from the web of S dualities involving the $I_4$ series SCFTs \cite{aw07}
\begin{align}
G_2 \with 8 \cdot {\bf 7}
&= A_1 \with {\bf 2}\oplus [II^*,C_5],
\label{Sd1}\\
B_3 \with 4 \cdot {\bf 8}\oplus 6 \cdot {\bf 7} 
&= C_2 \with 5 \cdot {\bf 4} \oplus [II^*,C_5],
\label{Sd2}\\
A_5 \with {\bf 21} \oplus \bar{\bf 21} \oplus {\bf 20} \oplus {\bf 6} \oplus \bar{\bf 6} 
&= A_4 \with {\bf 10} \oplus \bar{\bf 10} \oplus [II^*,C_5],
\label{Sd3}\\
C_2 \with 6 \cdot {\bf 5}
&= A_1 \with [III^*, C_3 A_1],
\label{Sd4}\\
C_2 \with 4 \cdot {\bf 4}\oplus 4 \cdot {\bf 5} 
&= A_1 \with 3 \cdot {\bf 2} \oplus [III^*, C_3 A_1],
\label{Sd5}\\
A_3 \with {\bf 10} \oplus \bar{\bf 10} \oplus 2 \cdot {\bf 4} \oplus 2 \cdot \bar{\bf 4}
&= A_2 \with {\bf 3} \oplus \bar{\bf 3}\oplus [III^*, C_3 A_1],
\label{Sd6}\\
A_2 \with {\bf 6} \oplus \bar{\bf 6} \oplus  {\bf 3} \oplus  \bar{\bf 3}
&= A_1 \with [IV^*,C_2 U_1].
\label{Sd7}
\end{align}
(We use Dynkin's notation for the simple Lie algebras, together with ``$U_1$" to stand for $\U(1)$.)
It is then easy to compute the dimension of the Higgs branches of these SCFTs.  For instance, from the first S duality, \eqref{Sd1}, the complex dimension of the Higgs branch on the left side is $8\cdot 7 - 2\cdot 14$:  there are $8\cdot 7$ complex scalars in the $8\cdot{\bf 7}$ half-hypermultiplets, of which $2\cdot 14$ are lifted by the vector multiplet in the $\bf 14$ (adjoint) of the $G_2$ gauge algebra by the $\cN=2$ Higgs mechanism (hyperk\"ahler quotient).  If the quaternionic dimension of the Higgs branch of the $[II^*,C_5]$ SCFT on the right side is $d_{\text{HB}}$, then the complex dimension of the Higgs branch on the right side is $2+2\cdot d_{\text{HB}} - 2\cdot 3$ by similar reasoning.  Equating the two sides gives $d_{\text{HB}} = 16$.  Other S-dualities involving the $[II^*,C_5]$ theory, such as \eqref{Sd2} and \eqref{Sd3}, give the same answer.  

In fact, since we know the explicit form of the left side Higgs branches from the hyperk\"ahler quotient construction, consistency among two or more such S duality relations should in principle suffice to determine the explicit structure of the Higgs branches of the SCFTs on the right side, along the lines of \cite{Gaiotto:2008nz}.  This method, however, is laborious; it may be easier to determine the Higgs branches by computing the associated ideal in the universal enveloping algebra of the flavor symmetry (i.e., the set of relations that a set of generators of the chiral ring satisfy) from the Hall-Littlewood index \cite{Beem:2014rza}.  In any case, we will not attempt to do these computations here.  

Similar reasoning gives the Higgs branch dimensions, $d_{\text{HB}}$, for the other SCFTs in the $I_4$ series, and their values are reported in table \ref{tab1}.  Note that the dimensions of these Higgs branches are not those of the minimal nilpotent orbits of the corresponding flavor algebras.  Thus, unlike the $I_1$ series, these Higgs branches do not coincide with centered 1-instanton moduli spaces.   

Also unlike the $I_1$ series, the $I_4$ series SCFTs all have non-trivial ECBs.   These can also be easily determined from the S-dualities \eqref{Sd1}--\eqref{Sd7} by matching the ECBs on both sides.

For example, from the first S duality, \eqref{Sd1}, the ECB fiber on the left side has quaternionic dimension 4 which transforms in the $\bf 8$ of the $C_4$ flavor symmetry.  On the right side, the $\bf 2$ half-hypermultiplet charged under the $A_1$ gauge factor does not contribute any ECB fiber, thus the $[II^*,C_5]$ SCFT must have an ECB fiber of quaternionic dimension at least 4.  Furthermore, the $A_1$ weakly gauges the $A_1$ subalgebra of the $C_5$ flavor symmetry with commutant $C_4$.  The ECB fiber of the SCFT must therefore have $h =5$ and transform in the $\bf 10$ of $C_5$ since upon weakly gauging the $A_1$ subalgebra, one of its quaternionic dimensions is lifted, giving the ECB of the left side.  It is easy to see that no other choice of symplectic representation of $C_5$ (potentially plus singlets) for the flavor action on the ECB fiber works.  This can furthermore be checked using other S-duality relations for the $[II^*,C_5]$ SCFT, such as those given in \eqref{Sd2} and \eqref{Sd3}.   Similar reasoning determines the ECB fiber of the $[III^*,C_3 A_1]$ SCFT as having $h =3$, transforming as the $({\bf 6},{\bf 1})$ under the flavor algebra.  

In the case of the $[IV^*,C_2 U_1]$ theory, the only known S-duality is the one shown in \eqref{Sd7}.   The left side has no ECB and has flavor symmetry $U_1\oplus U_1$, and on the right side the $A_1$ weakly gauges the index-2 $A_1 \subset C_2$ with commutant $U_1$ \cite{aw10}.  There are two possible consistent solutions for the ECB fiber of the $[IV^*,C_2 U_1]$ theory:  (1) the trivial one in which it is empty, and (2) one in which $h =2$ and it transforms as ${\bf 4}_0$ under the $C_2\oplus U_1$ flavor symmetry.  The second solution is consistent since the index-2 $A_1\subset C_2$ is the one under which the fundamental of $C_2$ decomposes as ${\bf 4}=2\cdot{\bf 2}$, and so the whole ECB fiber is lifted upon gauging the $A_1$.  Of these two solutions, only the second one is consistent with the behavior of the  $[IV^*,C_2 U_1]$ theory under RG flows.  In particular, upon turning on a mass adjointly breaking the flavor factor $C_2  \to C_1$ (with index of embedding 1, so that ${\bf 4} = {\bf 2} \oplus 2\cdot{\bf 1}$) it is known \cite{paper2} that the $[IV^*,C_2 U_1]$ theory flows to the $[I_0^*,C_1]$ theory which is the lagrangian $\cN=4$ $\SU(2)$ SYM theory.  Since this latter theory has a one-quaternionic-dimensional ECB fiber, its UV parent must have a non-empty ECB fiber as well.

These results for the $I_4$ series are shown in table \ref{tab1}.  It is curious that all the $I_4$ series SCFTs have ECB fibers which transform in the fundamental of a symplectic flavor symmetry factor, just as is always the case for gauge (lagrangian) theories, as we showed above.  This is possibly a result of these theories having a purely lagrangian dual.   But, as shown by the $I_1^*$ and $IV^*_{Q=1}$ series SCFTs (which have no known lagrangian duals), this pattern of ECB fibers does not hold for general SCFTs.  (The ECBs of these theories were determined in \cite{allm1602}.) 

\subsection{Chiral rings of SCFTs and coordinate ring of HBs and ECBs}
\label{sec2.3}

So far we have discussed the geometric aspects of the moduli space of $\cN=2$ SCFTs. The existence of these HBs and ECBs also puts constraints on the operator algebra of the associated SCFTs.  The possible $\cN=2$ superconformal multiplets containing scalar operators which can get vevs parametrizing the various branches give rise to a chiral ring in the SCFT operator product expansion.  This chiral ring may be identified with the coordinate ring of the moduli space of the SCFT.  We will show that a very simple assumption on the structure of the SCFT chiral ring reproduces the coordinate rings of the ECBs described above.

\subsubsection{$\cN$=2 SCFT chiral ring}

We first recall some facts about the 4d $\cN=2$ superconformal operator spectrum.  A primary field of a superconformal multiplet is characterized by its dimension $\D$, Lorentz spins $(j,\tj)$, $\SU(2)_R$ spin (``R-spin") $R$, and $\U(1)_R$-charge $r$.  The unitary, positive energy representations of the $\cN=2$ superconformal algebra, following \cite{Dobrev:1985qv, Dolan:2002zh} are summarized in table 6 of \cite{paper1}.

The most important representations for our purposes are the Lorentz scalar ``semi-chiral" $\cB_{R,r\,(0,0)}$ multiplets and their ``bi-chiral" $\cBh_R$ shortenings.  Vevs of the primaries of these multiplets can parameterize the moduli space of vacua of the SCFT.  For ease of notation, we will drop the Lorentz spin $(j,\tj)=(0,0)$ subscripts on the $\cB$ multiplets since we will only consider Lorentz scalars from now on.  $\cB_{R,r}$ have $\U(1)_R$ charge $r>1$ while $\cBh_R$ has $r=0$.  There are also Lorentz scalar $\cD_R$ multiplets which can be thought of as specializations (shortenings) of the $\cB_{R,r}$ scalar multiplets in the $r\to1$ limit.  Finally, the $R=0$ $\cB$ multiplets are anti-chiral, and are also called ``$\cE$"  multiplets, $\cB_{0,r} \equiv \cE_r$.  In all cases the dimension of the primaries of these multiplets is given by 
\begin{align}\label{chiraldims}
\D=2R+r .
\end{align}

By virtue of this relation, the Lorentz scalar $\cB_{R,r}$ multiplet primaries with maximal R-spin ($R_3=R$) form a chiral ring, as do the complex scalar $\cBh_R$ primaries with highest R-spin.  Also, it is easy to see that the product of a $\cBh$ maximal R-spin primary with a scalar $\cB$ multiplet maximal R-spin primary is another scalar $\cB$ primary.  

More precisely, pick an $\cN=1$ subalgebra of the $\cN=2$ algebra corresponding to a choice of Cartan $\U(1)_{R_3}\subset\SU(2)_R$, and denote the complex scalar primaries of the $\cBh_R$, $\cB_{R,r}$, and $\cE_r$ multiplets with $R_3=R$ by
\begin{align}\label{crnotn}
\cBh_R &\to q_R, &
\cB_{R,r} &\to m_{R,r}, &
\cE_r &\to \vf_r, &
\text{with}\ & R\in\{\tfrac12, 1, \tfrac32, 2, \ldots\} 
\ \text{and}\ r\ge1.
\end{align}
(The $m$, $\vf$ fields with $r=1$ are actually the complex scalar primaries of the $\cD_R$ multiplets: $\vf_1 \in \cD_0$ and $m_{R,1}\in\cD_R$.)
Then these complex fields satisfy the chiral ring relations
\begin{align}\label{chiralring}
q^a_R \, q^b_S &= C^{ab}_c q^c_{R{+}S},
&
q^a_R \, m^i_{S,s} &= C^{ai}_j m^j_{R{+}S\,,\,s} 
&
q^a_R \, \vf^\a_r &= C^{a\a}_i m^i_{R,r}
\notag\\
&&
m^i_{R,r} \, m^j_{S,s} &= C^{ij}_k m^k_{R{+}S\,,\,r{+}s}
&
m^i_{R,r} \, \vf^\a_s &= C^{i\a}_j m^j_{R\,,\,r{+}s}
\\
&&
&&
\vf^\a_r \, \vf^\b_s &= C^{\a\b}_\g \vf^\g_{r{+}s}
\notag
\end{align}
Here the $a$, $i$, $\a$ indices label fields in different multiplets with the same $(R,r)$ quantum numbers.  The $C^{\cdot\cdot}_\cdot$ are complex constants which determine the chiral ring up to the freedom to perform linear redefinitions of the fields within each $(R,r)$ sector.  

There are some constraints on the $C^{\cdot\cdot}_\cdot$'s which follow from physics.  First, $\cBh_0\ni q_0 \equiv 1$, is the identity and is assumed to be unique; we have dropped its trivial relations from \eqref{chiralring} by restricting the $R$ index to start at $\tfrac12$.  Second, $q^a_{1/2}\in\cBh_{1/2}$ and $\vf^\a_1\in\cD_0$ are scalars in free massless hypermultiplets\footnote{The other complex scalar field in the free hypermultiplet is the $R_3=-\tfrac12$ primary of the $\cBh_{1/2}$ multiplet;  because it is a free field, it is also part of the chiral ring.} and vector multiplets, respectively.  As such, they are free generators of the (commutative) chiral ring.  Third, the $\cBh_1$ multiplet containing the $q_1$ fields (with $R=1$ and $\D=2$) have a conserved current at the second level, so transform in the adjoint of the $\cN=2$ flavor symmetry algebra.  All other fields in the chiral ring can be organized into irreducible representations of the flavor group.  Since there is no chiral ring relation in \eqref{chiralring} of the form $q_1 \vf_r \sim \vf_r$, the $\vf^a_r$ are all flavor singlets.

It is possible, from their quantum numbers, to make an identification between the coordinate ring of the various branches and the complex scalar primaries of the $\cBh_R$, $\cB_{R,r}$, and $\cE_r$ multiplets generating the chiral ring relations \eqref{chiralring}:

\begin{itemize}

\item The $\boldsymbol{q^a_R}$ scalars carry no $\U(1)_R$ charge and have non-zero $\SU(2)_R$ spins, their vevs can be identified with the \textit{\textbf{Higgs branch}} complex coordinate ring.

\item The $\boldsymbol{\vf^\a_r}$ scalars can be identified with \textit{\textbf{Coulomb branch}} chiral ring operators, as they have zero R-spin but non-zero $\U(1)_R$ charge.

\item The $\boldsymbol{m_{R,r}}$ scalars carry instead both $\U(1)_R$ and $\SU(2)_R$ charge and can thus be identified with \textit{\textbf{mixed branch}} chiral ring operators.

\end{itemize}

We should note that the identification of all these chiral ring operators with Higgs, Coulomb, and mixed branch operators is conjectural \cite{Tachikawa:2013kta,Beem:2014zpa} in the sense that  it is possible that some or all of them may occur in the SCFT operator algebra but do not correspond to flat directions.  Conversely, however, if a $\vf_r$ field does develop a vev, then since it has $R=0$ the $\SU(2)_R$ symmetry remains unbroken.  Furthermore, by the Goldstone theorem, the Nambu-Goldstone boson of the spontaneously broken scale symmetry (the dilaton) must decouple in the IR. It will then be in a free $\cN=2$ supermultiplet whose scalars are $\SU(2)_R$ singlets, that is a vector multiplet.  A similar argument applies for the $q_R$ fields.

Finally, the chiral ring describes only the (or a) complex structure of the moduli space, but not its metric structure.  The hyperk\"ahler and special K\"ahler structures on the moduli space are encoded in other (singular) terms in the chiral primary OPEs.  We will not have anything to say about the metric structure of the moduli space in what follows.

\subsubsection{Moduli space coordinate ring for rank-1 SCFTs}

We take the CB to have planar topology, so its chiral ring will be freely generated by a single $\vf_r$.  The HB and ECB fiber are more difficult to describe.  We start by describing a simple example of the coordinate ring of a $\Z_2$ orbifold; this will prove useful later.

\paragraph{Coordinate ring of $V_d:=\C^d/\Z_2$.}   

This $d$-dimensional variety with an isolated singularity at the origin is defined as the orbifold of $\C^d$ under the equivalence $\bv \sim -\bv$ where $\bv = (v_1,\ldots,v_d) \in \C^d$.  Because of the $\Z_2$ identification, $\bv$ are not ``good" (i.e., globally defined) complex coordinates on $V_d$, but $z_{(ij)} := v_i v_j$ are good coordinates.  However, they are not independent since they satisfy the polynomial relations generating the ideal
\begin{align}\label{Z2ideal}
\cI = \vev{ \, z_{(ij)} z_{(k\ell)} {-} z_{(ik)} z_{(j\ell)}
\ ,\  \forall i,j,k,\ell\  }.
\end{align}
So we can think of $\{z_{(ij)}\}\in \C^n$ with $n= \frac12 d(d+1)$, and define $V_d= \C^n/\cI$.  Note that there are $\frac{1}{12}d^2(d^2-1)$ constraints generating \eqref{Z2ideal} even though dim$(V_d) = d$.  It is easy to see that one cannot eliminate any of the constraints in favor of the others because they are all of the same degree.  One can, however, check that $V_d$ is indeed $d$-dimensional by looking in the vicinity of a specific point, say one with $z_{11}\neq0$, and then systematically solving for all but $d-1$ of the other variables by dividing by $z_{11}$ as needed.

\bigskip

\noindent Let's now turn to the description of the chiral ring structure of the HB and ECB of the theories in table \ref{tab1}. We will only describe theories in the $I_4$, $I^*_1$ and $IV^*_{Q=1}$ series as the $I_1$ series has been already extensively discussed in the literature \cite{Gaiotto:2008nz}.

\paragraph{$\boldsymbol{ I_4}$ series.}

First, consider the moduli space of the $\cN=4$ $\SU(2)\simeq\SO(3)$ SYM theory.  This is the $[I_0^*,C_1]$ theory in the $I_4$ series in table \ref{tab1}.  Since it is a lagrangian theory, it is easy to fully describe its chiral ring.  It will turn out that the ECB's of the other (non-lagrangian) SCFTs in the $I_4$ series follow a similar pattern.

From an $\cN=1$ perspective, the $\cN=4$ R-symmetry splits as $\SO(6)_R \supset \SU(3)_F \times \U(1)_{r'}$ where the $\SU(3)_F$ is interpreted in the $\cN=1$ theory as a flavor symmetry.    In an $\cN=1$ lagrangian (gauge-variant) description, the moduli space is the space of vevs of the gauge adjoint complex scalars of three chiral multiplets.  Denote these fields as $Q^i_a$ with $a=1,2,3$ the gauge triplet index, and $i=1,2,3$ an $\SU(3)_F$ triplet index.  Then the $\cN=1$ superpotential is $\cW = \e_{ijk} Q^i_a Q^j_b Q^k_c \e_{abc}$, implying $F$-term constraints which are equivalent to
\begin{align}\label{N4Fterms}
Q_b^j Q^k_c = Q_c^j Q^k_b \quad \forall\ j,k,b,c . 
\end{align}
The $D$-term constraints are solved by forming all holomorphic gauge invariants, an algebraic basis of which are the ``mesons"
\begin{align}\label{N4Dterms}
\C^3 \ni M^{(ij)} := Q^i_a Q^j_a .
\end{align}
Then the moduli space is given in terms of the mesons by
\begin{align}\label{N4V}
V_{\cN{=}4} = \C^3 /\cJ, \quad
\cJ = \vev{\ M^{(ij)}M^{(k\ell)} {-} M^{(ik)}M^{(j\ell)}
\ ,\ \forall\ i,j,k,\ell\ },
\end{align}
where the relations generating the ideal $\cJ$ follow from \eqref{N4Fterms}.  We recognize this variety as $V_{\cN{=}4} = V_3 = \C^3 /\Z_2$.

Now, from the $\cN=2$ perspective, the $\cN=4$ R-symmetry splits instead as $\SO(6)_R \supset \SO(4) \times \SO(2) \simeq \SU(2)_F \times \SU(2)_R \times \U(1)_r$, where $\SU(2)_F\simeq C_1$ is the $\cN=2$ flavor symmetry shown in table \ref{tab1}.  The $Q^i_a$ of the $\cN=1$ description for $i=1,2$ then transform as a doublet under the diagonal $\SU(2) \subset \SU(2)_F \times \SU(2)_R$ and are not charged under the $\U(1)_r$, while $Q^3_a$ is a flavor- and $R$-singlet of charge 1 under $\U(1)_r$.  
Thus the meson fields fall into $\cN=2$ supermultiplets as
\ytableausetup{boxsize=1.2mm}
\begin{align}\label{}
M^{(ij)} &= q_1^{\yd{2}} 
& &\in \cBh_1 & & &
\text{with}\ &\ R=1,\ r=0,\ F=1,
& 
\nonumber\\
M^{(i3)} &= m_{\frac12,1}^{\yd{1}} 
& &\in \cB_{\frac12,1} & & &
\text{with}\ &\ R=\tfrac12,\ r=1,\ F=\tfrac12,
& 
\\
M^{(33)} &= \vf_2 
& &\in \cE_2 & & &
\text{with}\ &\ R=0,\ r=2,\ F=0,
& 
\nonumber
\end{align}
where the Young diagram superscripts denote the $\SU(2)_F$ representation, and we are using the notation of equation \eqref{crnotn}.  Then the \eqref{N4V} constraints imply the (leading) chiral ring \eqref{chiralring} relations
\begin{align}\label{N4cr1}
q_1^{\yd{2}} \, q_1^{\yd{2}} 
&\sim q_2^{\yd{4}},
&
q_1^{\yd{2}} \, m_{\frac12,1}^{\yd{1}} 
&\sim m_{\frac32,1}^{\yd{3}} 
&
q_1^{\yd{2}} \, \vf_2^\bullet 
&\sim m_{1,2}^{\yd{2}}
\\  \label{N4cr2}
&&
m_{\frac12,1}^{\yd{1}} \, m_{\frac12,1}^{\yd{1}} 
&\sim m_{1,2}^{\yd{2}}
&
m_{\frac12,1}^{\yd{1}} \, \vf_2^\bullet 
&\sim m_{\frac12,3}^{\yd{1}}
\\  \label{N4cr3}
&&
&&
\vf_2^\bullet \, \vf_2^\bullet
&\sim \vf_4^\bullet .
\end{align}
\ytableausetup{boxsize=1.5mm}%
Since $\yd2 \otimes_S \yd2 = \yd4 \oplus \bullet$, the first relation of \eqref{N4cr1} reflects the constraint that the singlet hypermultiplet does not appear in the HB chiral ring.  (This is the simplest instance of the Joseph ideal relation defining the minimal nilpotent orbit for $\SU(2)_F$.)    Similarly, since $\yd2 \otimes \yd1 = \yd3 \oplus \yd1$, the second relation of \eqref{N4cr1} reflects that constraint that 
\ytableausetup{boxsize=1.2mm}%
$m_{\frac32,1}^{\yd1}$ does not appear in the ECB chiral ring.
\ytableausetup{boxsize=1.5mm}%
Since $\yd2 \otimes \bullet = \yd2 = \yd1\otimes_S\yd1$, there is no constraint in the third chiral ring relation in \eqref{N4cr1} or first relation in \eqref{N4cr2} involving the vanishing of ECB mutliplets, but there is a constraint that the \emph{same} ECB field, 
\ytableausetup{boxsize=1.2mm}%
$m_{1,2}^{\yd{2}}$, appears on the right side of both relations in order to be compatible with \eqref{N4V}.

We will now extend this analysis to theories with $C_n\simeq \Sp(n)$ flavor groups with $n>1$.  The cases with $n=2,3,5$ appear among the non-lagrangian SCFTs of the $I_4$ series in table \ref{tab1}.   The relevant $C_n$ representation theory is summarized in
\ytableausetup{boxsize=1.5mm}%
\begin{align}\label{axsa}
\yd2 \otimes_S \yd2 &= \yd4 \oplus 
\left[\, \raisebox{3pt}{$\yd{2,2}$}\oplus 
\raisebox{3pt}{$\yd{1,1}$} \oplus \bullet \,\right],
\\ \label{axf}
\yd2 \otimes \yd1 &= \yd3 \oplus 
\left[\, \raisebox{3pt}{$\yd{2,1}$} \oplus \yd1 \, \right],
\\ \label{fxsf}
\yd1 \otimes_S \yd1 &= \yd2 = \yd2 \otimes \bullet .
\end{align}
Then the chiral ring relations analogous to \eqref{N4cr1}--\eqref{N4cr3} but with only the $\U(1)_r$ charges changed,
\ytableausetup{boxsize=1.2mm}%
\begin{align}\label{Cncr1}
q_1^{\yd{2}} \, q_1^{\yd{2}} 
&\sim q_2^{\yd{4}},
&
q_1^{\yd{2}} \, m_{\frac12,\frac{r}2}^{\yd{1}} 
&\sim m_{\frac32,\frac{r}2}^{\yd{3}} 
&
q_1^{\yd{2}} \, \vf_r^\bullet 
&\sim m_{1,r}^{\yd{2}}
\\  \label{Cncr2}
&&
m_{\frac12,\frac{r}2}^{\yd{1}} \, 
m_{\frac12,\frac{r}2}^{\yd{1}} 
&\sim m_{1,r}^{\yd{2}}
&
m_{\frac12,\frac{r}2}^{\yd{1}} \, \vf_r^\bullet 
&\sim m_{\frac12,\frac{3r}2}^{\yd{1}}
\\  \label{Cncr3}
&&
&&
\vf_r^\bullet \, \vf_r^\bullet
&\sim \vf_{2r}^\bullet ,
\end{align}
imply the constraints:  (1) the Higgs branch fields carrying the representations in brackets in \eqref{axsa} do not appear on the right side of the first relation in \eqref{Cncr1}; (2) the mixed branch fields carrying the representations in brackets in \eqref{axf} do not appear on the right side of the second relation in \eqref{Cncr1}; and the (3) mixed branch fields, $m^{\yd2}_{1,r}$, appearing in \eqref{Cncr1} and \eqref{Cncr2} are the same.  Constraint (1) is the Joseph ideal constraint describing the minimal nilpotent $C_n$ orbit.  This is $2n$-complex-dimensional, so has the correct dimension to describe the ECB fiber over the origin of the CB for the $I_4$ series SCFTs.   In fact, these three constraints suffice to describe the whole ECB.  To see this, call $q_1^{\yd2} := M^{(ij)}$, $m_{\frac12,\frac{r}2}^{\yd1} := M^{(i0)}$, and $\vf_r^\bullet = M^{(00)}$, for $i,j,k,\ell \in\{1,\ldots,2n\}$.  Then 
\begin{align}\label{CnECBconstr}
&&\text{constraint (1)} &&&\Rightarrow&
M^{(ij)} M^{(k\ell)} &= M^{(ik)} M^{(j\ell)} ,&&
\nonumber\\
&&\text{constraint (2)} &&&\Rightarrow&
M^{(ij)} M^{(k0)} &= M^{(ik)} M^{(j0)} ,&&
\\
&&\text{constraint (3)} &&&\Rightarrow&
M^{(ij)} M^{(00)} &= M^{(i0)} M^{(j0)} .&&
\nonumber
\end{align} 
Comparing to \eqref{Z2ideal}, we see that these relations describe the coordinate ring of $V_{2n+1} := \C^{2n+1}/\Z_2$.  This fits nicely with the description of the ECBs of the $I_4$ series given in table \ref{tab1}.   In particular, dim$_\C(\text{ECB}) = 1+2h  = 1+2n$ for $n=5$ for the $[II^*,C_5]$ theory, for $n=3$ for the $[III^*,C_3A_1]$ theory, and for $n=2$ for the $[IV^*,C_2U_1]$ theory.   Thus in each case the ECB is a $\Z_2$ orbifold.

The above chiral ring does not capture the HBs of the $I_4$ series SCFTs with a $C_n$ flavor factor, which as we have seen have complex dimensions $2d_{\text{HB}}$ given by $32$, $16$ and $8$ for $n=5$, 3, and 2, respectively.  They presumably arise in the chiral ring by loosening constraint (1) in \eqref{CnECBconstr}, e.g., by allowing Higgs branch fields $q_2^{\cdots}$ to appear on the right side of the first chiral ring relation in \eqref{Cncr1} carrying one or more of the $C_n$ irreps in the Joseph ideal (those in brackets in \eqref{axsa}).  

It is tempting to try to identify these HBs as nilpotent $C_n$ orbits.\footnote{The nilpotent orbits of $C_n$ can be described in terms of partitions of integers and their associated Young diagrams as follows \cite{cm93}.  Label nilpotent orbit $\cO_\bd$ by a partition $\bd \equiv [d_1, \cdots, d_r]$ of $2n$ such that each odd $d_i$ occurs with even multiplicity.  The associated Young diagram is the one with $d_i$ boxes in its $i$th row, where the $d_i$ are put in non-increasing order. Define the transpose partition by $\bd^t \equiv  [p_1,\cdots, p_s]$ with $p_i := |\{j|d_j \geq i\}|$, so that the $p_i$ are the lengths of the rows of the transpose Young diagram of $\bd$, which is the Young diagram with rows and columns exchanged.
The complex dimension of a given nilpotent orbit $\cO_\bd$ is given by dim$(\cO_\bd) = n (2n+1) -\frac{1}{2} (\S + P)$, where $\S := \sum_i p_i^2$, and $P := |\{j|d_j \text{ is odd}\}|$.   The closure of one nilpotent orbit contains another, $\bar{\cO_\bd} \supset \bar{\cO_{\bd'}}$, if and only if  $\sum_{i=1}^k d_i \geq \sum_{i=1}^k d'_i$ for all $k$.  Under this partial ordering, the minimal non-zero nilpotent $C_n$ orbit is $\bd=[2,1^{2n-2}]$ and has dimension $2n$.  It is contained in the closure of all other nilpotent $C_n$ orbits.}  
It is easy to check that for $C_n$, there is a unique orbit of dimension $8(n-1)$, and it is special.\footnote{This orbit is the orbit identified as  $\bd=[4,2,1^{2n-6}]$ which, for $n=2$, degenerates to $\bd=[4]$. A $C_n$ partition $\bd$ is \emph{special} if $\bd^t$ is also a $C_n$ partition, which is equivalent to it having an even number of even entries. Special orbits are those which can be described by ``primitive ideals" constructed from irreducible representations of $C_n$ in the ring of polynomials in dim$(C_n)$ variables.}  For $n=5$, $3$, and $2$, these give the right dimensions to be the HBs of the $I_4$ series CFTs.  So it is natural to conjecture that these are, in fact, the HBs of these theories.


\paragraph{$\boldsymbol{ I^*_1}$ and $\boldsymbol{IV^*_{Q=1}}$ series.} The ECB of the $[II^*,A_3\rtimes\Z_2]$ CFT was determined in \cite{allm1602} to have fiber of complex dimension 8, transforming as ${\bf 4}\oplus\bar{\bf4}$ of $A_3$, while its HB was determined in \cite{Chacaltana:2016shw} to have complex dimension 18.  Though there is an 8-dimensional $A_3$ nilpotent orbit, its coordinate ring does not fit with that of the ECB fiber over generic points on the CB to give simple chiral ring relations such as \eqref{Cncr1}--\eqref{Cncr3} in the $I_4$ series.  Furthermore, there is no 18-dimensional nilpotent orbit for $A_3$.

The ECBs of the other $I^*_1$ and $IV^*_{Q{=}1}$ series CFTs were also determined in \cite{allm1602} and are shown in table \ref{tab1}.  However the HBs of the $[III^*,A_1U_1\rtimes\Z_2]$ and $[II^*,A_2\rtimes\Z_2]$ theories could not be determined.  The chiral rings of the ECBs (which include the HBs) of the $[IV^*,U_1]$ and $[III^*,U_1\rtimes\Z_2]$ are very simple and were described in \cite{allm1602}.


\section{Central charges from the twisted ECB partition function}
\label{sec3}

We now turn to the computation of the conformal and current algebra central charges of rank-1 $\cN=2$ SCFTs.  We follow and slightly extend the method of Shapere and Tachikawa \cite{st08} to compute these central charges from the low energy data on the CB.  The dimension of the ECB and the action of the flavor group on it turn out to be important inputs to this computation.  We again use S dual relations of some of the rank-1 SCFTs to weakly coupled $\cN$=2 gauge theories as independent checks.  

\subsection{Topologically twisted ECB partition function}
\label{sec3.1}

The $a$ and $c$ central charges of the 4d conformal algebra are certain coefficients in OPEs of energy-momentum tensors, and the $k$ central charges appear in the OPEs of flavor currents.  In case the flavor algebra $\ff = \oplus_a \ff_a$ is a sum of simple or $\U(1)$ factors, then each factor will have a separate $k_a$ central charge.   These OPE coefficients are special because they appear in the scale anomaly in the presence of a background metric and background gauge fields for $\ff$ as coefficients multiplying certain scalar densities of the background fields. $\cN=2$ superconformal symmetry relates the scale anomaly to 't Hooft anomalies for the $\U(1)_R\oplus\SU(2)_R\oplus_a\ff_a$ global symmetry, with the result that in the presence of a background metric and background gauge fields for the global symmetries, the conservation of the $\U(1)_R$ current is broken by terms proportional to the central charges times topological densities formed from the background fields. Shapere and Tachikawa \cite{st08} were able to use these results to relate the computation of the $a$ and $c$ central charges of SCFTs in flat space to the $\U(1)_R$ anomalies of the topologically twisted gauge theories obtained from the initial $\cN=2$ SCFTs. A slight generalization of this method is needed to compute the central charges in our cases. But first we will briefly review their method involving the topologically twisted CB partition function \cite{st08}. For a more detailed treatment we refer to the original literature \cite{st08,w88,w95}

Background metric and gauge fields describing an arbitrary smooth oriented 4-fold $M$ with $F$-bundle (where $F$ is the flavor symmetry group with Lie algebra $\ff$) generally break $\cN=2$ supersymmetry.   However, if one chooses the background $\SU(2)_R$ field strength proportional to the self-dual part of the background curvature, a topologically twisted sector of the theory --- sensitive only to topological invariants of the background fields --- is still protected by a supersymmetry \cite{w88}.  The result \cite{w95,st08} is that the partition function of the twisted theory on $M$ with an $F$-bundle carries $\U(1)_R$ charge\footnote{Note that we are using a normalization of the $\U(1)_R$ charge such that $R(u) = \D(u)$.  This differs from that used in \cite{st08} by a factor of two.}
\begin{align}\label{DRxsn}
\D R = (2a-c)\cdot \chi + \frac32 c \cdot \s - \frac12 \sum_a k_a \cdot n_a,
\end{align}
where $\chi$ and $\s$ are the Euler number and signature of $M$ and $n_a$ are the instanton numbers of the $F$-bundle.

We normalize the instanton numbers in the usual way so that they run over all integers as we vary the 4-fold $M$ and the $F$-bundles with $F$ simply connected.  Since we are working on general $M$ the partition function may be sensitive to the global form of $F$.  We will use this in section \ref{sec3.4} below to put some constraints on what the global form of the flavor symmetry can be in certain SCFTs.

Then \eqref{DRxsn} corresponds to the standard normalizations of the central charges where for $n_V$ free vector multiplets and $n_H$ free hypermultiplets
\begin{align}\label{normack}
24 a &= 5 n_V + n_H, &
12 c &= 2 n_V + n_H, &
k_a &= T_a({\bf 2n_H}).
\end{align}
Thus, in this case
\begin{align}\label{freeDRxsn}
\D R_\text{free} = \frac14 n_V \cdot \chi 
+ \left(\frac14 n_V + \frac18 n_H \right)\cdot\s 
- \frac12 \sum_a T_a({\bf 2n_H}) \cdot n_a.
\end{align}
Here ${\bf 2n_H}$ is the (reducible) representation of $\ff$ under which the $2n_H$ half-hypermultiplets transform.  $T_a({\bf 2n_H})$ is the quadratic index of ${\bf 2n_H}$ with respect to the $\ff_a$ factor.\footnote{If ${\bf 2n_H}$ decomposes into irreps of $\oplus_{a=1}^L \ff_a$ according to ${\bf 2n_H} = \oplus_\a (\br_{\a1}\otimes\br_{\a2}\otimes\cdots\otimes\br_{\a L})$, then $T_a({\bf 2n_H}) = \sum_\a \left(\prod_{b\neq a} r_{\a b} \right) T(\br_{\a a})$. The quadratic index for a simple factor is proportional to the sum of the squared-lengths of weights in ${\bf 2n_H}$, $T({\bf 2n_H}) := (1/\text{rank}\ff)\sum_\l (\l,\l)$, where the weights are normalized so that the long roots of $\ff$ have length-squared 2.  This is the normalization for which $T(\bn)=1$ for $\SU(n)$.}    In case $n_H=0$, there is no contribution from the last term in \eqref{freeDRxsn}, so we adopt the convention that $T(``{\bf 0}"):=0$.

Upon flowing to the IR on the Coulomb branch, the partition function of the twisted theory is given by the path integral of the low energy Lagrangian \cite{w95}
\begin{align}\label{Ztwist}
Z=\int [dV] [dH]\ \cA^\ch\ \cB^\s \ \textstyle{\prod_a} \cC_a^{n_a} 
\ e^{S_\text{lR}[V,H]} .
\end{align}
The path integral is over the $n_V$ (IR free) massless neutral vector multiplet fields and $n_H$ massless neutral hypermultiplet fields (if any) on the ECB.  This includes an ordinary integral over the 0-modes (constant modes) of the vector multiplet scalars, $u$. It can be shown that $\cA$, $\cB$, and the $\cC_a$ can only depend holomorphically on $u$, the masses, and the chiral relevant or marginal deformation parameters.  They can have zeros or poles only at singularities of the ECB where additional states (i.e., beyond those described by $V$ and $H$) become massless.

Thus from \eqref{Ztwist} the total $\U(1)_R$ charge of the partition function is evaluated at a generic (i.e., non-singular) point on the CB to be
\begin{align}\label{DRABC}
\D R = \left(R(\cA) + \frac14 n_V \right)\cdot \chi 
+ \left(R(\cB) + \frac14 n_V + \frac18 n_H \right)\cdot\s 
+ \sum_a \left(R(\cC_a) - \frac12 T_a({\bf 2n_H}) \right)\cdot n_a,
\end{align}
where we have used \eqref{freeDRxsn} to evaluate the contribution from the $[dV][dH]$ measure.  Comparing this to \eqref{DRxsn} for arbitrary $(\ch,\s,n_a)$ gives
\begin{align}\label{acki}
24 a &= 5 n_V + n_H + 12 R(\cA) + 8 R(\cB) ,
\nonumber\\
12 c &= 2 n_V + n_H + 8 R(\cB) ,
\\
k_a &= T_a({\bf 2n_H}) - 2 R(\cC_a) .
\nonumber
\end{align}

These are our key equations, relating the central charges to the low energy data $n_V$, $n_H$, ${\bf 2n_H}$, $\cA$, $\cB$, $\cC_a$.  Since $n_V$ is just the complex dimension of the CB, in the rank 1 case we are examining here,
\begin{align}\label{}
n_V = 1.
\end{align}
Since $n_H$ is the number of massless netural hypermultiplets at a generic point on the CB, it is the quaternionic dimension of the Higgs fiber of the ECB,
\begin{align}\label{}
n_H = h .
\end{align}
It remains to determine the dimensions of $\cA$, $\cB$, and the $\cC_a$.

Topological invariance implies \cite{w95} that $\cA$, $\cB$, and $\cC_a$ depend holomorphically on $u$.   It can be shown that $\cA^\ch$ transforms as a holomorphic modular form of weight $-\ch/2$ under the EM duality group on the CB \cite{w95}, and this fixes $\cA$ to be \cite{mw97, mm97, mm98} $\cA = \a \det(\del u^i/\del a^j)^{1/2}$.  Here $a^j$ are the special coordinates on the CB which have scaling dimension, and thus $\U(1)_R$ charge, 1.  The prefactor $\a$ is $u$-independent.  In the conformal case it can only depend on constants which are all dimensionless, so $R(\a)=0$.  In the rank-1 case, therefore, we have
\begin{align}\label{RofA}
R(\cA) = \frac{\D-1}2 
\end{align}
where $\D:=\D(u)$ is the scaling dimension of the global complex coordinate on the CB.

When the twisted theory is put on a smooth spin 4-manifold, $\cB^\s$ and $\cC_a^{n_a}$ are single-valued functions on the CB.\footnote{For non-spin 4-manifolds the $\cB^\s$ measure factor may be multi-valued on the CB \cite{w95}.}    Since $\cB^\s$ is holomorphic in $u$ and for smooth spin 4-manifolds $\s\in16\Z$ \cite{rohlin52,teichner92}, we see that $\cB^{16}$ must be a single-valued holomorphic function of $u$.   Likewise, the instanton numbers $n_a\in\Z$ (at least for simply-connected flavor groups; see section \ref{sec3.4}), so $\cC_a$ must be a single-valued holomorphic function of $u$.

We discuss the determination of $R(\cB)$ and $R(\cC_a)$ in the next two subsections.

\subsection{Conformal algebra central charges}
\label{sec3.2}

In the rank-1 case, by holomorphy and scale invariance, we must have that $\cB^{16}= \b u^b$ for some integer $b$ and complex constant $\b$, so $R(\cB) = b \D/16$.  Comparing to \eqref{acki} gives in the rank-1 case the following integer associated to SCFTs,
\begin{align}\label{binvt}
b := 16 \frac{R(\cB)}{\D} = 2\frac{12c-2-h}{\D} \in \Z.
\end{align}
We determine the integer $b$ by the same line of argument used in \cite{st08}.  The only slight difference is that we show we can do so without assuming weak coupling asymptotically far on the CB.  

The strategy is to consider deforming away from conformality by turning on some $\cN=2$ preserving relevant operators.  Such a deformation does not lift the Coulomb branch, though it does generically lift the ECB fibers.  The $\cB^{16}$ function will be deformed to one which can only have a zero or a pole in $u$ at some singular points, $u=u_i$, on the CB, i.e., points where there are additional massless states.  The scaling dimension of $\cB^{16}$ near these points then reflects the contribution of these additional degrees of freedom.  Also, the large-$u$ asymptotics of $\cB^{16}$ are not changed by the deformation since relevant deformations have arbitrarily small effect at large $u$; see \cite{paper1} for a discussion.  Since we are assuming that for arbitrary relevant deformation the CB is simply the complex $u$-plane (i.e., does not have a more complicated topology), it then follows that we can determine the total degree of $\cB^{16}$ in $u$ at $u=0$ for the undeformed theory by summing the degrees of $\cB^{16}$ at each $u=u_i$ for the deformed theory.

For generic deformation, we know from \cite{paper1} that the CFT singularity at the origin of the CB splits into undeformable singularities at points $u_i$, $i=1,\ldots,Z$.\footnote{The $u_i$'s are the zeros in the $u$-plane --- \emph{not} counted with multiplicity --- of the discriminant of the SW curve, see again \cite{paper1}.}  Each one of these singularities is associated to a SCFT; that is, the additional massless states at $u_i$ together with any generic massless neutral vector or hypermultiplets form an IR CFT, denoted as CFT$_i$.  Each CFT$_i$ is rank 1 with a CB coordinate $u-u_i$ of dimension $\D_i$, conformal central charges $a_i$ and $c_i$, flavor algebra $\ff_i$ with central charge $k_i$, and an ECB fiber of quaternionic dimension $h_i$ which transforms under $\ff_i$ in representation $\br_i$.

Then applying the first two equations in \eqref{acki} to CFT$_i$, we solve for $R(\cA_i)$ and $R(\cB_i)$ as
\begin{align}\label{AiBi}
12 R(\cA_i) &= 24 a_i -12 c_i -3,\nonumber\\
8 R(\cB_i) &= 12 c_i -2-h_i .
\end{align}
Note that \eqref{RofA} applied to CFT$_i$ implies
\begin{align}\label{2a-c}
24 a_i - 12 c_i &= 3 (2\D_i - 1),
\end{align}
a relation thus predicted \cite{st08} for all $\cN=2$ SCFTs, and which can be checked directly for conformal gauge theories \cite{aw07}.  Given a value for $R(\cB_i)$, we deduce that $\cB_i^{16} =\b_i (u-u_i)^{16 R(\cB_i)/\D_i}+ \ldots$ where $\b_i$ is a holomorphic function of $\bm$ and the dots are subleading terms as $u\to u_i$.  Recalling that $\cB_i^{16}$ must be single-valued in $u$ implies that
\begin{align}\label{c-quant}
16 \frac{R(\cB_i)}{\D_i} =
2 \frac{12 c_i -2-h_i}{\D_i} := b_i
\in\Z.
\end{align}

Call the $\cB^{16}$ function for the deformed theory $\cB^{16}(u,\bm)$.  We recover the $u$-dependence of $\cB^{16}(u)$ for the original (undeformed) CFT as $\lim_{\bm\to0} \cB^{16}(u,\bm)$, since the CB dimension does not change as a function of the $\bm$.  (Note that in general we have to take a limit, and not simply set $\bm=0$ since $\cB^{16}(u,\bm)$ may diverge at $\bm=0$ if the ECB dimension jumps at $\bm=0$, i.e., when there is an ECB fiber for $\bm=0$ but none (or a smaller one) for $\bm\neq0$.)

Since there can be no zeros or poles in $\cB^{16}$ other than those due to the CFT$_i$, we learn that $\cB^{16} = \b u^b$ with 
\begin{align}\label{bsumrule}
b =  \sum_{i=1}^Z b_i .
\end{align}
Then \eqref{binvt} and \eqref{c-quant} gives the final answer for $R(\cB)$ for the original CFT.  Plugging into \eqref{acki} gives
\begin{align}\label{ac-formula}
24 a &= 5 + h  + 6 (\D-1) + \D \sum_{i=1}^Z \frac{12 c_i -2-h_i}{\D_i},
\nonumber\\
12 c &= 2 + h  + \D \sum_{i=1}^Z \frac{12 c_i -2-h_i}{\D_i} .
\end{align}
This expresses the conformal central charges of the SCFT in question in terms of its CB coordinate dimension, $\D$, the quaternionic dimension, $h$, of its ECB fiber, and the analogous data $(\D_i,h_i,c_i)$ for the IR SCFTs at each of the singularities that appear upon generically deforming it.

For generic values of the deformation parameters the set of CFT$_i$'s is given by the deformation pattern in table \ref{tab1}.  The CFT$_i$'s that can appear are thus those undeformable CFTs with Kodaira singularity of type $I_n$, $n\in\{1,2,4\}$ or $I^*_1$ or $IV^*_{Q=1}$.  These CFTs, discussed at length in \cite{paper1,paper2}, have the following properties: 
\begin{itemize}
\item[$I_n$:]
An undeformable $I_n$ singularity corresponds to an IR free $\U(1)$ gauge theory with a single charge $\sqrt{n}$ massless hypermultiplet.  It thus has the field content of one free vector multiplet and one free hypermultiplet, a CB field of dimension 1, no higgs branches, and a $\U(1)$ flavor symmetry under which the free half-hypermultiplets have charges $1\oplus(-1)$  (in an arbitrary normalization of the $\U(1)$ flavor current).  Thus for these theories, $\D=1$, $24a=6$, $12c=3$, $\ff=\U(1)$, $k=2$, and $h=0$, 
where we used \eqref{normack}.  In particular, all undeformable $I_n$ singularities contribute $b=1$ in \eqref{c-quant}.  These are independent of $n$ since $n$ can be absorbed in the normalization of the generators of the $\U(1)$ gauge group, which has no physical significance.
\item[$I_1^*$:]
The frozen $I_1^*$ singularity that appears in the $I^*_1$ series arises from the IR free $\SU(2)$ gauge theory with a single massless \emph{half}-hypermultiplet in the $\bf 4$ irrep (a.k.a.\ the spin-3/2 irrep).  It thus has the field content of 3 free vector multiplets and 2 free hypermultiplets, a CB field of dimension 2, no higgs branches, and no flavor symmetry.  Thus for this theory, $\D=2$, $24a=17$, $12c=8$, $\ff=\varnothing$, $h=0$, and so the frozen $I_1^*$ singularity contributes $b=3$ in \eqref{c-quant}.
\item[$IV^*_{Q{=}1}$:]
Finally, the frozen $IV^*_{Q{=}1}$ singularity arises from a (hypothetical) interacting CFT with a CB field of dimension 3 and no flavor symmetry, implying that $\D=3$, $24a=12c+15$, $c:=c'$, $\ff=\varnothing$, $h=0$, and that it contributes $b=2(12c'-2)/3$ in \eqref{c-quant}.  In this case, the value, $c'$, of its central charge must be determined from other arguments.
\end{itemize}

It is now straightforward to apply \eqref{ac-formula} to the regular rank 1 deformation patterns listed in table \ref{tab1}, reproducing the values of the $a$ and $c$ conformal central charges listed there.  In the cases where the rank 1 SCFTs in table \ref{tab1} can be related to weakly coupled lagrangian SCFTs either directly (for the deformations of the $I^*_0$ singularity) or using S-dualities (for the unshaded deformations of the $II^*$, $III^*$, and $IV^*$ singularities in table \ref{tab1}), the conformal central charges can be independently calculated  \cite{as07,aw07}.  The agreement between these two methods was already noted in \cite{aw10}.  Note that there is only agreement between these two methods once the contributions of the neutral hypermultiplets on the ECB are correctly accounted for.

In the case of the $IV^*_{Q{=}1}$ series, the values of the central charges cannot be computed from first principles. The ones shown in table \ref{tab1} followed from the \emph{assumption} that the $[III^*,U_1\rtimes\Z_2]$ theory has $\cN{=}3$ supersymmetry, and was discussed in \cite{allm1602}.  
Note that according to the S-fold arguments of \cite{gr1512,at1602}, an $\cN=3$ $[III^*,U_1\rtimes\Z_2]$ SCFT with these central charges is expected to exist.

\subsection{Current algebra central charges}
\label{sec3.3}

We will only compute the flavor central charges, $k$, for simple (and therefore nonabelian) factors of the flavor symmetry.  Central charges for $\U(1)$ factors of flavor groups are difficult to determine using these techniques because of the possibility of them mixing under RG flows with the low energy global electric and magnetic $\U(1)$'s on the CB \cite{st08}.  Furthermore, these $\U(1)$ central charges are only defined relative to a choice of normalization of the $\U(1)$ generators.  Thus a $k_{\U(1)}$ central charge needs to specified together with the $\U(1)$ flavor charge of a BPS particle in the theory in order to have meaning.   For these reasons, we do not list the $\U(1)$ flavor central charges in table \ref{tab1}:  for those with a ``?" in the central charge entry, we are unable to compute it, while for those with a ``*", it can be calculated relative to a conventional normalization.  For example, for the undeformable IR-free $I_1$, $I_2$, and $I_4$ theories appearing in table \ref{tab1}, $k_{\U(1)}=2$ in the normalization where the $\U(1)$ flavor charges of the free hypermultiplets are $\pm 1$.  Also $k_{\U(1)}$ can be calculated relative to a given normalization for the $[IV^*,U_1\rtimes\Z_2]$ and $[III^*,U_1]$ theories in table \ref{tab1} since they are $\cN=3$ SCFTs and so the $\U(1)$ flavor symmetry of their $\cN=2$ deformations is part of the $\cN=3$ $\U(3)$ R-symmetry, implying its central charge is proportional to the $a=c$ central charge. 

For simplicity, assume that the flavor symmetry, $\ff$, is simple; it is easy to generalize the following argument to semi-simple $\ff=\oplus_a\ff_a$.  Then there is a single $\cC^n$ factor in the twisted partition function on the ECB of our CFT, and, by holomorphy and scale invariance, 
\begin{align}\label{Cnuen}
\cC^n = \g u^{-en},
\end{align}
for some integer $e$ and complex constant $\g$.  (The sign of $e$ is for later convenience.)  Comparing to \eqref{acki} gives in the rank-1 case the following integer associated to SCFTs,
\begin{align}\label{einvt}
e := -\frac{R(\cC)}{\D} = \frac{k-T({\bf 2h})}{2\D} \in \Z ,
\end{align}
where $\D=\D(u)$ is the scaling dimension of the CB parameter vanishing at the conformal vacuum of the CFT.

Because of the difficulty with computing $\U(1)$ central charges, we can not use the strategy of the last section of turning on a generic mass deformation, since under such a deformation the low energy flavor group is entirely broken to $\U(1)$ factors.   Instead, we will use special mass deformations which leave some nonabelian subalgebra of the SCFT flavor symmetry unbroken.  

Suppose that under one such special mass deformation, $\bm$, our $[K,\ff]$ SCFT  (here $K$ is the Kodaira type and $\ff$ is the flavor symmetry) deforms to $Y$ distinct singularities as
\begin{align}\label{}
[K,\ff\,] \xrightarrow{\bm} \left\{ [K_1, \ff_1\oplus\uf_1], \ldots, [K_Y,\ff_Y\oplus\uf_Y]
\right\} .
\end{align}
Here we have separated out the semi-simple part, $\ff_j$, from the abelian factors, $\uf_j$, of the flavor algebra for each singularity.  We will now focus on just one of these singularities, say the $i$th one, corresponding to a $[K_i,\ff_i\oplus\uf_i]$ CFT.  

Put the topologically twisted theory in a background of $n_i$ instantons only in $\ff_i\subset\ff$.  This corresponds to a total $n$-instanton background for the original $\ff$ flavor symmetry where 
\begin{align}\label{nnidi}
n = n_i d_i \qquad\qquad\text{(no summation)}, 
\end{align}
and the $d_i$ are the Dynkin indices of embedding $\ff_i \hookrightarrow \ff$.

Suppose the flavor central charge of the $\ff_i$ factor at the $[K_i,\ff_i]$ singularity is $k_i$, the quaternionic dimension of its ECB fiber is $h_i$, and, as usual, $\D_i$ is the scaling dimension of the $u-u_i$ CB parameter there.  Then this CFT$_i$ contributes a factor $\cC_i$ with R-charge (dimension) 
\begin{align}\label{ki}
R(\cC_i) = (T({\bf 2h_i}) - k_i)/2
\end{align} 
by \eqref{acki}.   Since 
\begin{align}\label{Ciniexp}
\cC_i^{n_i} \sim (u-u_i)^{-e_i n_i} + \ldots,
\qquad\qquad e_i \in \Z,
\end{align}
for some integer $e_i$, where the dots represent subleading terms, we find from \eqref{ki}
\begin{align}\label{Diei}
e_i = \frac{k_i-T({\bf 2h_i})}{2\D_i} \in \Z.
\end{align} 

Then the $u$-dependence of the $\cC^n$ measure factor of the original CFT is given by turning off the mass deformation: $\cC^n = \lim_{\bm\to0} \cC_i^{n_i}$.  Comparing \eqref{Cnuen}, \eqref{Ciniexp}, and \eqref{nnidi} gives
\begin{align}\label{}
d_i e= e_i,
\end{align}
which implies, from \eqref{einvt} and \eqref{Diei}, that
\begin{align}\label{kform}
k &= \frac{\D}{d_i \D_i} \left(k_i - T\left({\bf 2h_i}\right)\right) + T({\bf 2h}) ,
\qquad\qquad
\text{for each simple $\ff_i$.}
\end{align}
Note that because any simple factor $\ff_i$ can be chosen, and also many different mass deformations, $\bm$, can be chosen, this formula for $k$ is highly over-determined.

We will now apply this formula to compute the flavor central charges of the theories shown in table \ref{tab1}.  In preparation, first observe that the $2h$ free half-hypermultiplets on the ECB fiber of the original $[K,\ff]$ CFT will transform in some representation of the subalgebra $\ff_i$, ${\bf 2h} = {\bf 2r_i} \oplus (2s_i \cdot {\bf 1})$.  Here we have separated off all the $2s_i$ singlets, so $2h=2r_i + 2s_i$.  (The symplectic nature of the representation, discussed in section \ref{sec2}, ensures that $2r_i$ is even.)  It is important to note that $r_i$ is \emph{not} necessarily the quaternionic dimension, $h_i$, of the ECB fiber of the $[K_i,\ff_i\oplus\uf_i]$ CFT.  The reasons for this are two-fold: $r_i$ may be smaller than $h_i$ to the extent that some of the $s_i$ singlet hypermultiplets might not be lifted because they are neutral under the $\uf_i$ abelian flavor factors; and $r_i$ may be larger than $h_i$ to the extent that some irreducible summands of $\bf 2 r_i$ may be charged under the abelian $\uf_i$ flavor factors, and so be lifted.
This implies, in particular, that the Dynkin index of embedding, $d_i := T({\bf 2r_i})/T({\bf 2h})$, appearing in \eqref{kform} is not simply expressible in terms of $T({\bf 2h_i})$.

\paragraph{\emph{I}$\bf_1$ series.}

Consider the $[II^*,E_8]$ CFT with CB parameter of dimension $\D=6$, deformed by turning on masses implementing the minimal adjoint flavor breaking $E_8 \to A_7\oplus U_1$.  From the SW curve for this theory, it is straightforward to compute that the singularity splits as
\begin{align}\label{}
[II^*, E_8 ] \to \{ [I_8, A_7\oplus U_1], [I_1,U_1], [I_1,U_1]\}.
\end{align}
The $[I_8,A_7\oplus U_1]$ CFT is an IR free $\U(1)$ gauge theory with 8 massless charge-1 hypermultiplets.  As such, the dimension of its CB parameter is $\D_1=1$, its $A_7$ flavor central charge is $k_1 = T({\bf8}\oplus{\bf\bar8}) = 2$.  Also, the Dynkin index of embedding of $A_7 \subset E_8$ is $d_1=1$.  Finally, neither the $[II^*,E_8]$ nor the $[I_8,A_7\oplus U_1]$ CFTs have an ECB fiber, so the $T({\bf2h})$ and $T({\bf2h_1})$ terms in \eqref{kform} should be dropped.  The result is $k = (6\cdot 2)/(1\cdot 1) = 12$, giving the result in \ref{tab1}, and in agreement (of course) with the value computed in \cite{at07,aw07,st08}.

It is interesting to perform similar computations with other flavor breakings, and check that the same answer for the $E_8$ flavor central charge results.   As an example, consider the minimal adjoint breaking $E_8 \to E_7\oplus U_1$, for which the singularity splits as
\begin{align}\label{}
[II^*, E_8 ] \to \{ [III^*, E_7], [I_1,U_1]\}.
\end{align}
This gives $\D=6$, $\D_1=4$, $d_1=1$, and $T({\bf2h})=T({\bf2h_1})=0$ to give $k = (6\cdot k_1)/(4\cdot 1)$, correctly giving the ratio of the $[II^*, E_8 ]$ and $[III^*, E_7]$ flavor central charges.  There are many more such checks that can be done, all giving the unique results for the $I_1$ series shown in \ref{tab1}. 

\paragraph{\emph{I}$\bf_4$ series.}

Similar computations for the $I_4$ series are more involved as they now involve non-trivial ECB fibers.  We illustrate with the $[II^*,C_5]$ CFT, which has $\D=6$, $h=5$, and $T({\bf10})=1$.
First consider the minimal adjoint flavor breaking $C_5 \to A_4 \oplus U_1$.  From the SW curve found in \cite{paper2}, one learns that under this deformation the singularity splits as
\begin{align}\label{}
[II^*,C_5] \to \{ [I_5, A_4\oplus U_1], [I_4,U_1],[I_1,U_1]\}.
\end{align}
The CFT associated to the first singularity in this list is the IR free $\U(1)$ gauge theory with 5 massless charge-1 hypermultiplets, and so has $\D_1=1$, $k_1 = T({\bf 5}\oplus{\bf\bar5})=2$, $h_1=0$.  Also $d_1=2$ for $A_4 \subset C_5$.  This then gives $k = \frac{6}{2\cdot1}(2-0)+1 = 7$.

Next, consider the ``opposite" minimal adjoint flavor breaking $C_5 \to U_1 \oplus C_4$ under which the singularity splits as
\begin{align}\label{}
[II^*,C_5] \to \{ [I_3^*, C_4],[I_1,U_1]\}.
\end{align}
The CFT associated to the first singularity in this list is the IR free $\SU(2)$ gauge theory with 8 massless adjoint half-hypermultiplets, as discussed in some detail in section 5.1 of \cite{paper2}.  This IR free theory has $\D_1=2$, $k_1 = 3\cdot T({\bf 8})=3$, and $h_1=4$ so $T({\bf2h_1})=T({\bf8})=1$.  The ECB fiber dimension follows from the discussion in appendix \ref{sec2.1.2}, since the $\bf 3$ of $\SU(2)$ has a single zero weight.  Also, $d_1=1$ for $C_4 \subset C_5$.  This then again gives $k = \frac{6}{1\cdot2}(3-1)+1 = 7$.

As a final example, consider the $C_5 \to U_1\oplus A_1 \oplus C_3$ minimal adjoint flavor breaking under which the singularity splits as
\begin{align}\label{}
[II^*,C_5] \to \{ [III^*, A_1\oplus C_3], [I_1,U_1]\}.
\end{align}
We can now compute $k$ for the $C_5$ theory by looking at either the $A_1$ or the $C_3$ simple flavor factors of the $III^*$ CFT on the right.  Call the $A_1$ and $C_3$ flavor central charges $k_1$ and $k_3$, respectively, and note that $d_1=2$ for $A_1\subset C_5$ while $d_3=1$ for $C_3\subset C_5$.  Then applying \eqref{kform} to the $A_1$ factor imples that $k = \frac{6}{2\cdot4}(k_1-0)+1$ where we have used that the ECB of the $III^*$ theory is neutral under the $A_1$ factor.  Similarly, for the $C_3$ factor, we find $k= \frac{6}{1\cdot4}(k_3-1)+1$.  For $k=7$, these then imply $(k_3,k_1)=(5,8)$, as shown in table \ref{tab1}.  

Again, as for the $I_1$ series, there are many more such checks that can be done, all giving the unique results for the $I_4$ series shown in \ref{tab1}.  Furthermore, an independent check is the computation of the flavor central charge using the S-duality equivalences, such as \eqref{Sd1}--\eqref{Sd7}, for the $I_4$ series of CFTs.  These were computed in \cite{aw07,aw10}, and agree with \eqref{kform}.

\paragraph{Other series.}

For the other series shown in table \ref{tab1} (as well as for some other theories not shown in table \ref{tab1}) the computations of the flavor central charges using \eqref{kform} were discussed in some detail in \cite{allm1602}.\footnote{Note that there is an error in the formula for $k$ in the published version of \cite{allm1602}; it is corrected in the arXiv version of that paper.} 

\subsection{Constraints from bounds and integrality conditions on central charges}
\label{sec3.4}

As a further check of the consistency of the picture thus far presented, we will check in this section that the values of the central charges computed using the techniques outlined above are consistent with known (and some conjectural) bounds in the literature.  We conclude this section by arguing that a careful analysis of the single-valuedness of the measure in the twisted partition function \eqref{Ztwist} can shed light on the existence of discrete anomalies in gauging the flavor symmetries of certain non-lagrangian theories with no known S-dual description.  This is a remarkable result as it relies on purely non-perturbative methods.

\paragraph{Bounds.}

The Hofman-Maldacena bounds \cite{Hofman:2008ar, Komargodski:2016gci, Hofman:2016awc} on $a/c$ for unitary $\cN=2$ SCFTs are  $1 \le 2a/c \le 5/2$.  The lower bound is saturated by theories of free hypermultiplets, and the upper bound by free vector multiplets.  These bounds are satisfied by all the theories in table \ref{tab1}.  Indeed, this can be shown to follow with mild assumptions from the topologically twisted ECB partition function formalism \cite{st08}.  The lowest value of $2a/c$ ($\sim 1.53$) is given by the $I_1$-series $[II^*,E_8]$ CFT, and the highest value ($2.2$) is given by the $IV^*_{Q{=}1}$-series $[IV^*,\varnothing]$ CFT.

Also, note that not only does $a$ decrease along RG flows within each series (in agreement with the $a$ theorem \cite{Komargodski:2011vj,Komargodski:2011xv}), but also $2a/c$ increases.  For weakly couped lagrangian SCFTs this behavior follows because the mass terms which generate the flow lead to the integrating out of at least as many hypermultiplets as the number of vector multiplets which are lifted by adjoint Higgsing on the CB (i.e., tuning the CB vev).  This means that the ratio of the number of vector multiplets to the number of hypermultiplets is non-decreasing along the flow, implying from \eqref{normack} that $2a/c$ is non-decreasing.  It is interesting that this pattern also seems to hold for non-lagrangian theories.

The $c$ bound $12 c\ge  22/5$ for interacting $\cN=2$ SCFTs \cite{Liendo:2015ofa} is also satisfied by all our theories and is saturated by the $I_1$-series $[II,\varnothing]$ CFT.

Analytic bounds on flavor central charges have been obtained from demanding positivity of certain SCFT OPE coefficients coming from unitarity, giving \cite{b+13,Lemos:2015orc}
\begin{align}\label{}
\frac{h^\vee}{k} &\le \frac12 + \frac12 \cdot \frac{|\ff|}{12c},
& &\text{and} & 
\frac{h^\vee}{k} &\le 3 - \frac{36}{5\cdot 12c-22} \cdot \frac{|\ff|}{12c},
\end{align}
where $h^\vee$ is the dual Coxeter number of $\ff$.
As discussed in \cite{b+13,Lemos:2015orc}, these bounds are saturated by the $I_1$-series CFTs except for the $[III,A_1]$ theory.  It is easy to check that none of the other theories in table \ref{tab1} saturate these bounds.  Beem et.\ al.\ \cite{b+13} also find $c$-independent unitarity bounds for $k$ which depend on the flavor algebra $\ff$, and are summarized in table 3 of their paper.  These are also saturated for the same subset of the $I_1$-series mentioned above.  For $\ff=C_n$, this bound is $k\ge n+2$, and it is interesting to note that this bound is saturated also for the $C_n$ flavor factors of the $I_4$-series CFTs of table \ref{tab1}.

\paragraph{Relations.}

For SCFTs related by S-dualities to lagrangian SCFTs, $a$ and $c$ are related by $24a-12c=3\sum_i (2\D(u_i)-1)$ where $\{u_i\}$ are a basis of good CB complex coordinates of definite scaling dimensions \cite{aw07}.  As mentioned above, this also follows from the topologically twisted ECB partition function formalism \cite{st08}, and so necessarily holds for all the theories in table \ref{tab1}.  It should be pointed out, however, that this relation is known to fail for SCFTs which involve a discrete gauging, which indicates that the  topologically twisted ECB partition function formalism as described here must be modified for these kinds of theories \cite{am1604}.
 
Another lagrangian result is that the quaternionic dimension of the Higgs branch is given by $24c-24a=n_H-n_V=d_\text{HB}$, as follows from the $\cN=2$ Higgs mechanism.  This relation has also been conjectured to extend to non-lagrangian SCFTs which have 3d mirror duals \cite{Xie:2013jc}.  It is easy to check that this relation is violated in a few low-$a$ CFTs in table \ref{tab1}:  the $I_1$-series $[II,\varnothing]$ theory has $24c-24a = 1/5$, and the $IV^*_{Q{=}1}$-series $[IV^*,\varnothing]$ theory has $24c-24a=-5/2$, while both have $d_\text{HB}=0$.  In all other cases, though, the formula works.  If one assumed that the formula also works for the two theories in table \ref{tab1} for which the HB dimension could not be determined by other arguments, it would predict that $d_\text{HB}=3$ for the $I_1^*$-series $[III^*,A_1U_1\rtimes\Z_2]$ CFT, and $d_\text{HB}=5$ for the $IV^*_{Q{=}1}$-series $[II^*,A_2\rtimes\Z_2]$ CFT.

\paragraph{Integrality conditions.}

As noted earlier in \eqref{binvt} and \eqref{einvt}, single-valuedness of the ECB twisted partition function measure factors $\cB^\s \cC^n$ implies that the combinations of central charges appearing in \eqref{binvt} and \eqref{einvt} must be integers:
\begin{align}\label{beinvts}
b &:= 2\frac{12c-2-h}{\D} \in \Z, &\text{and}& &
e &:= \frac{k-T({\bf 2h})}{2\D} \in \Z .
\end{align}  
This followed because $\cB^\s \cC^n \sim u^{-en+b(\s/16)}$ and because the signature of smooth spin 4-manifolds $M$ is divisible by 16, $\s\in16\Z$, and in our normalization of the instanton number $n\in\Z$ for $F$-bundles with simply-connected flavor group $F$.

The values of $b$ are shown in table \ref{tab1}, and they are all, indeed, integers.  In fact, they are all even except for the theories in the $IV^*_{Q{=}1}$ series which have odd $b$.   Note that, conjecturally, there may be other rank-1 SCFTs corresponding to the other possible deformed rank-1 CB geometries listed in table 1 of \cite{paper1}.  The central charges of these theories cannot be completely determined by the techniques of this paper since upon deformation they all flow to frozen non-lagrangian CFTs (like the $IV^*_{Q{=}1}$ theory shown in table \ref{tab1}).  However, assuming frozen CFTs have no ECBs, the above integrality condition constrains their central charges to be $12c \in (\D/2)\Z + 2$, where $\D=3,4,6$ for the hypothetical frozen $IV^*$, $III^*$, $II^*$ CFTs, respectively.  The $24a-12c=3(2\D-1)$ relation fixes $a$ in terms of $c$, and the strongest bound on $\Z$ then comes from the $2a/c \le 5/2$ bound which then implies $\Z\ge2$ for $\D=3$ and $\D=4$ (in the latter case $\Z=2$ saturates the bound), and $\Z\ge3$ for $\D=6$.  Similar integrality conditions and bounds can be obtained for the non-lagrangian interpretations of the $I^*_n$ frozen singularities as well. 

The values for $e$ are also shown in table \ref{tab1}, and they are integers for all theories except for certain ones in the $I_2$ and $I_4$ series, where they are half-odd-integral.  This means that for those theories, in certain 1-instanton backgrounds for the flavor symmetry the $\cC^n$ factor in the twisted partition function is double-valued, and as a result the partition function is not well-defined, as its sign is ambiguous.   

What is the physical interpretation of this sign ambiguity? We argue that it precisely reflects the existence of a $\Z_2$ obstruction \cite{w82} to gauging the flavor symmetry $\ff$ of these theories. In fact such $\Z_2$ obstruction appears as a sign ambiguity in the partition function of a gauge theory \cite{w82} which is also the effect of having half-odd $e$ in the twisted partition function on the ECB.  In lagrangian theories this obstruction occurs when there are Weyl fermions in a (possibly reducible) symplectic representation $\bf2r$ of $\ff$ with odd quadratic index $T({\bf2r})$.  (These only occur for Lie groups with non-vanishing $\pi_4$, which are ones whose Lie algebras have simple $C_N \simeq \Sp(2N)$ factors.)   In non-lagrangian theories the obstruction can be tracked by 't Hooft anomaly matching by adding an additional decoupled Weyl fermion transforming in an appropriate symplectic representation of $\ff$. This can also be seen from the behavior of the partition function. In fact a decoupled half-hypermultiplet, $H'$, in an appropriate symplectic representation, $\bf 2r$, of $\ff$ precisely restores the single-valuedness of the ECB partition function.  This is not because the free half-hypermultiplet makes $e$ integral, in fact $e$ remains unchanged: the half-hypermultiplet contributes a factor of $T({\bf 2r})$ in \eqref{beinvts} but at the same time $k$ also increases by precisely the same amount so the half-hypermultiplet contributions to $e$ in cancel. Instead, upon traversing a cycle, $\g$, in the CB enclosing the conformal vacuum, the measure of the twisted partition function \eqref{Ztwist} of the CFT still gains a minus sign.  But with the addition of the free half-hypermultiplet, $H'$, there is another contribution, $[dH']$, to the measure in \eqref{Ztwist} which contributes a cancelling minus sign upon traversing $\g$ because of the $\Z_2$ twist of its ECB fiber, discussed in appendix \ref{sec2.1.2} below.

As mentioned above the only simple Lie algebras with symplectic representations with odd quadratic index are the symplectic ones, $C_N\simeq \Sp(2N)$, so only CFTs with these simple flavor factors can have a $\Z_2$ obstruction.    Of the topmost CFTs in each of the series in table \ref{tab1}, only those of the $I_2$ and $I_4$ series have such flavor symmetry factors.  The $[II^*,C_5]$ CFT of the $I_4$ series indeed has the $\Z_2$ obstruction, as deduced in \cite{aw07} from $\cN=2$ S-dualities such as \eqref{Sd1}--\eqref{Sd3}.   The $[I^*_0,C_1]$ theory of the $I_2$ series is the $\cN=4$ $\SU(2)$ gauge theory, which, from the $\cN=2$ perspective, has a $C_1$ doublet of $\SU(2)$ gauge-triplet half-hypermultiplets which therefore give a $\Z_2$ obstruction to gauging the $C_1$ flavor symmetry.   Any potential $\Z_2$ obstructions for the lower CFTs in each series then follow by flowing from the top theory.  For instance, since all the $C_n$ flavor factors in the $I_4$ series are realized upon suitable mass deformations as subgroups of the $C_5$ flavor symmetry of the $[II^*,C_5]$ CFT with Dynkin index of embedding 1, it follows by an 't Hooft anomaly matching argument that they all have $\Z_2$ obstructions to being gauged.  This same line of argument applied to the other series then gives the $\Z_2$ obstructions recorded in table \ref{tab1}.  

The integrality of $e$ followed from the integrality of the possible instanton numbers of the background flavor bundles on the 4-fold $M$.  We normalized the instanton numbers in the standard way\footnote{E.g., by defining $n:= \frac{1}{8\pi^2} \int_M (\Omega \overset{\wedge}{,} \Omega)$, where $\Omega$ is the 2-form background $\ff$-valued field strength on the euclidean 4-fold $M$, and $(\cdot,\cdot)$ is the Killing form on $\ff$ normalized so that the length-squared of long roots is 2.} so that instanton numbers are integers for arbitrary $F$-bundles over $M$, where $F$ is the simply connected compact Lie group with Lie algebra $\ff$.   But if $F$ is not simply connected, then, in this normalization, there can be $M$ for which there are $F$-bundles with fractional instanton number.  In particular, it is shown in \cite{Dijkgraaf:1989pz} that SO$(3)=$ SU$(2)/\Z_2$ has instantons with charges in $\Z/4$.   This extends to PSp$(2N)=$ Sp$(2N)/\Z_2$, e.g.\ by embedding the SO$(3)\simeq$ PSp$(2)$ bundle over $\C\P^2$ constructed in \cite{Dijkgraaf:1989pz} into PSp$(2N)$ for $N>1$.  From this we can deduce that the global form of the Lie group of the $C_N$ flavor Lie algebras appearing in the $I_4$ series CFTs must be the simply connected form Sp$(2N)$ and not PSp$(2N)$.  This is because if it were PSp$(2N)$ then for the $\cC^n\sim u^{-en}$ measure factor to be single-valued, since $n\in\Z/4$, we would need to have $e\in4\Z$.  In fact, as we discussed in the last paragraph, for the $I_4$ series CFTs, the measure factor must in fact be double-valued, but this would still imply $e\in2\Z$, in contradiction with the half-odd-integral values for $e$ computed for these theories.  Similar arguments can be used to constrain the possible global forms of the flavor symmetry groups of the other CFTs appearing in table \ref{tab1}. 


\section{Summary and open questions}
\label{sec4}

\paragraph{Summary of results.}  This paper, and the other papers in this series, implement a program to classify possible ``rank 1, planar" $\cN=2$ SCFTs, i.e., those whose CBs, as complex manifolds, are isomorphic to $\C$, the complex plane.
\begin{itemize}
\item In \cite{paper1,paper2} we classified and constructed all possible planar special K\"ahler geometries which can be consistently interpreted as the CB of either an SCFT or an IR-free theory.  We gave evidence for a ``safely irrelevant conjecture":  there are no $\cN{=}2$-preserving dangerously irrelevant RG flows.  This allowed a classification of the possible distinct rank-1 planar CB geometries in terms of families labeled by a scale-invariant geometry (the ``UV singularity") and its complex deformations.  The generic deformation gives a geometry with a characteristic set of singularities (the ``IR singularities").  

\item In most cases, the IR singularities are all of Kodaira types $I_n$ or $I_n^*$, and have simple interpretations as ``undeformable" IR-free gauge theories.  In this case there are further physical consistency conditions restricting the allowed CB geometries coming from imposing the Dirac quantization condition on the low energy theory (not only for generic values of the deformation parameters, but also for all special values where some of the IR singularities merge).  In a few cases the generically deformed CB geometries have IR singularities of types $IV^*$ or $III^*$ which could be consistently interpreted as new ``frozen" $\cN=2$ SCFTs. This was also discussed and analyzed in \cite{paper1,paper2}.

\item Furthermore, as discussed in some detail in section 5.3 of \cite{paper2}, geometries with $I^*_n$ or $I_n$ IR singularities can also be consistently interpreted as weakly-gauged non-lagrangian ``frozen" CFTs if one is willing to posit the existence of a class of rank-0 $\cN=2$ SCFTs.  Since there is no independent evidence for the existence of such rank-0 SCFTs, and since there are few additional physical constraints (beyond $\cN=2$ supersymmetry) that can be put on the resulting rank-1 geometries, we simply listed the $I^*_n$-series of geometries in \cite{paper1,paper2} but gave no discussion of the properties of their (potential) associated SCFTs beyond the discussion in section 5.3 of \cite{paper2}.

\item For the remaining CB geometries, we then turned to deducing properties of their associated SCFTs.  As discussed in section 4.4 of \cite{paper1}, the flavor symmetry can not be directly deduced from the CB geometry.  Instead, only a discrete group can be deduced which must include the Weyl group of the flavor symmetry.  In \cite{allm1602} we determined the complete set of allowed flavor symmetry groups that are consistent with our CB geometries and also consisent under RG flows.  There results a long list of possible consistent distinct rank-1 SCFTs.  Many of those SCFTs were found in \cite{am1604} as the result of gauging certain discrete symmetries of other (known) rank-1 SCFTs and IR-free theories.

\item Finally, this paper outlined how to compute (or at least constrain) the central charges of SCFTs from their CB geometries following \cite{st08}.  To do so, we found we needed to understand the properties of ECBs. The geometrical and algebraic structure of ECBs and HBs are also extensively discussed in the present manuscript.
\end{itemize}

What is the final result on the set of possible planar, rank-1 $\cN=2$ SCFTs?  As discussed in \cite{paper1,paper2}, the possible CB geometries can be organized into series by what kind of singularities appear in a generic RG flow to the IR.  The series are $I_1$, $I_2$, $I_4$, $I_0^*$, $I_1^*$, $I_2^*$, $I_3^*$, $IV^*$, $III^*$, and $II^*$.  (The last is not a ``series": it is simply a conjectural frozen $[II^*,\varnothing]$ SCFT.)   Which of these series correspond to actual SCFTs? We summarize our results in table \ref{tab1}:
\begin{itemize}
\item[1)] The $I_1$ series (or the ``maximal deformation" series) all correspond to SCFTs and have been found as flows from asymptotically free $\cN=2$ gauge theories \cite{Argyres:1995jj,Argyres:1995xn} or are related by S-dualities to weakly-coupled $\cN=2$ gauge SCFTs \cite{Minahan:1996fg,Minahan:1996cj,as07,aw07}.  

\item[2)] The $I_2$ series flows from a lagrangian ($\cN=4$) theory, and so exist \cite{sw2}.

\item[3)] The $I_4$ series, like the $I_1$ series, have been found via S-dualities \cite{aw07,aw10,Chacaltana:2014nya}.  They are characterized by having symplectic flavor symmetries.

An interesting note is that the $I_4$ and the $I_2$ series both contain versions of the CB geometry of the $\cN=4$ $\SU(2)$ gauge theory.   We discussed the relation between these two geometries in section 3.4 of \cite{paper2}, and it was further explored in \cite{am1604}. 

\item[4)] The $I_1^*$ series in table \ref{tab1} is more inherently strongly coupled: its $[II^*,A_3\rtimes\Z_2]$ member is constructed using class $\cS$ techniques \cite{Chacaltana:2016shw}, but does not seem to be S-dual to any gauge theory, while its $[IV^*,U_1]$ member is an $\cN=3$ SCFT predicted by S-fold arguments \cite{gr1512,nt1602,at1602}.
\end{itemize}
These four series all have the feature that they flow to free theories in the IR upon (generic) deformation.  In this sense, they are natural generalizations of the original $[I_0^*,D_4]$ and $[I_0^*,C_1]$ Seiberg-Witten CB solutions \cite{sw2}.  
\begin{itemize}
\item[5)] In table \ref{tab1} we have also included the $IV^*_{Q{=}1}$ series, which flows to a (conjectural) frozen interacting $[IV^*,\varnothing]$ SCFT.  The evidence for this is two-fold. These are the only CB geometries that can accommodate the $\cN=3$ $[III^*,U_1\rtimes\Z_2]$ SCFTs predicted by S-fold arguments \cite{gr1512,nt1602,at1602}. Furthermore there is at least one known W-algebra, namely the W(2,7) in \cite{Blumenhagen:1990jv}, which could be consistently interpreted as 2d chiral algebra associated to the frozen $IV^*_{Q{=}1}$ singularity\footnote{We thank Madalena Lemos for pointing this out to us.}.
\end{itemize}

Can other series of SCFTs which flow to frozen interacting SCFTs in the IR also exist?  There are certainly other possible CB geometries ending in frozen $IV^*$, $III^*$, or $II^*$ singularities.  Furthermore, as explained in \cite{paper2} and mentioned above, it is possible to re-interpret $I_n$ (for $n\ge1$) and $I_n^*$ (for $n\ge0$) singularities as exotic rank-0 frozen SCFTs coupled to free vector multiplets. If we allowed for such interpretation a plethora of new geometries would be consistent, yet there seems to be no independent evidence for the existence of such rank-0 theories at present.

Another possible way of realizing various IR ``frozen" singularities is as IR-free theories with an appropriate \emph{discrete} global symmetry gauged.  When the discrete symmetry acts on the CB, gauging it changes the CB geometry.  There is strong evidence for the existence of discretely gauged versions of many of the CFTs in the 5 series shown in table \ref{tab1}, and, interestingly, they provide examples of all but 3 of the possible CB geometries shown in table 1 of \cite{paper1} and table 1 of \cite{paper2}.  This is explained in detail in \cite{am1604}.  

Thus, a conservative {\bf conjecture} is: 
\begin{quote}\emph{The only planar, rank-1 $\cN=2$ SCFTs are those in table \ref{tab1} together with their discrete gaugings listed in \cite{am1604}.}
\end{quote} 

\paragraph{Further directions.}  In addition to some more technical questions mentioned in the conclusions to \cite{paper1, paper2}, some questions raised by those papers and this one include:
\begin{itemize}
\item Can the Shapere-Tachikawa central charge calculus \cite{st08} be refined to also compute the flavor central charges for $\U(1)$ factors despite the possibility of mixing with accidental $\U(1)$'s in the IR? 

\item Can the Shapere-Tachikawa central charge calculus \cite{st08} be modified to apply to discretely-gauged SCFTs \cite{am1604}?

\item How do the techniques of \cite{paper1,paper2} and this paper  generalize to non-planar CB geometries, and is there any independent evidence for the existence of such SCFTs? (Some results in this direction appear in \cite{paperIIc}.)

\item Our method can be generalized to any rank CB, but computationally the problem becomes considerably more complicated already at rank 2 \cite{rank-2}.
Computational complexity aside, it is an interesting question whether the set of physical conditions outlined in \cite{paper1,paper2} and this paper would, in principle, enable a complete classification of SCFT CB geometries at ranks 2 and higher.

\item Is there an intrinsic characterization (i.e., just in terms of the geometrical structures on the CB) of the physically allowed CB deformation patterns?
\end{itemize}
Finally, there is the question of connections to other work on SCFTs.  We have pointed out at various points in \cite{paper1, paper2, allm1602} and in this paper some connections to class $\cS$ constructions, to F-theory or S-fold constructions, to SCFT index computations, and to the analytic and numerical bootstrap program.  It would be interesting to also clarify the connections of our program to the results coming from geometric engineering of $\cN=2$ SCFTs, to 3d $\cN=4$ SCFTs via compactification, and to work on BPS quivers. Furthermore it worth pointing out the results in \cite{Hellerman:2017sur} which could shed light in how to directly connect the results obtained through our study of the CB geometry with those obtained studying the algebra of operators at the conformal vacuum.


\begin{acknowledgments}

It is a pleasure to thank P. Esposito, M. Lemos, C. Long, A. Shapere, Y. Tachikawa, R. Wijewardhana and J. Wittig, for helpful comments and discussions.  This work was supported in part by DOE grant DE-SC0011784.  MM was also partially supported by NSF grant PHY-1151392.

\end{acknowledgments}

\appendix

\section{Gauge theory ECBs} \label{sec2.1.2}

In this section we collect some technical arguments to show the results quoted \ref{ECBs} which, to make the reading easier, we also report here:

\begin{itemize}

\item[i)] In an $\cN=2$ gauge conformal field theory, ECBs occur whenever there are hypermultiplets in a representation $R$ of the gauge group which has zero weights (e.g., $\SU(2)$ integer spin representations). It can be shown that such representations are necessarily orthogonal, though the converse is not true.

\item[ii)] In $\cN=2$ gauge theories with hypermultiplets transforming in, generally reducible, representations $R$ of the gauge group, the most general flavor symmetry group is a direct sum of unitary, orthogonal and symplectic factors:
\begin{align}\label{ffdecomp}
\ff = \left[\oplus_i \U(\ell_i)\right] \oplus 
\left[\oplus_j\SO(m_j)\right] \oplus 
\left[\oplus_k \Sp(2n_k)\right] ;
\end{align}
ECB's can only occur in the theories with symplectic flavor factors.

\item[iii)] The ECB hyperk\"ahler factor transforms as a direct sum of fundamental representations of (some subset of) the flavor symmetry $\ff$ symplectic factors.

\item[iv)] Generally the ECB fiber, as we approach a singularity in the CB, degenerates into a cone $\H^h/\sim_\s$, where $\sigma$ is a triholomorphic isometry of $\H^h$ which fixes the origin and commutes with the flavor group. For lagrangian SCFTs, the twist $\s$ is always in the $\Z_2$  center of the appropriate symplectic flavor factor.

\end{itemize}

The general coupling of a half-hypermultiplet, $Q$, in a (generally reducible) representation, $R$, of the gauge group to the vector multiplet scalar, $\Phi$, appears in the Lagrangian in the term 
\begin{align}\label{supotl}
\int d^2\th\, \Phi^A Q_I J^{IJ} (T^A_R)_J^K Q_K 
\end{align}
in an $\cN=1$ superfield notation where $Q$ stands for the chiral superfield of the half-hyperplet, $\Phi^A$ is the $\cN=1$ chiral multiplet in the $\cN=2$ gauge multiplet (so $A$ is a gauge adjoint index), $T^A_R$ are the gauge generators in the $R$ representation (so $I,J,\ldots$ are $R$ indices), and $J^{IJ}$ is a symplectic pairing on $R$ inherited from the symplectic form on the $\H^{\text{dim}(R)/2}$ moduli space of hypermultiplet scalars in the zero gauge coupling limit.  This means that $J$ is a non-degenerate antisymmetric invariant tensor intertwining $R$ and its conjugate, $JT^A_R = -(T^A_R)^T J$, and it only exists if $R$ is a symplectic (a.k.a.\  pseudoreal) representation.  $R$ must therefore be decomposable into a direct sum of symplectic, or pairs of orthogonal, or conjugate pairs of complex representations,
\begin{align}\label{Rdecomp}
R = \left[\oplus_i \ell_i \left(C_i\oplus \bar C_i\right)\right] \oplus
\left[\oplus_j m_j S_j\right] \oplus
\left[\oplus_k 2 n_k O_k\right] ,
\end{align}
where $C_i$, $S_j$, and $O_k$ are all distinct complex, symplectic, and orthogonal irreducible representations, and $\ell_i$, $m_j$, and $2 n_k$ are their multiplicities.\footnote{While  $m_j$ can be either even or odd, $\Z_2$ anomaly cancellation \cite{w82} imposes further restrictions on odd $m_j$; see, e.g., section 4.2 of \cite{paper1}.}  The flavor symmetry is then 
\begin{align}\label{ffdecomp2}
\ff = \left[\oplus_i \U(\ell_i)\right] \oplus 
\left[\oplus_j\SO(m_j)\right] \oplus 
\left[\oplus_k \Sp(2n_k)\right] ;
\end{align}
see, e.g., \cite{mmw13}.    

On the CB where $\Phi^A$ can be taken in a complexified Cartan subalgebra of the gauge algebra, their associated gauge generators $\{T^i_R\}$, $i=1,\ldots,r$, can be diagonalized.  That is, there is a basis $\{Q_{\l,\a}\}$ of $R$ such that $T^i_R Q_{\l,\a}  = \l^i Q_{\l,\a}$.  Here $\l^i$ are the weights of $R$, and the $\a$ index labels any possible multiplicities of each weight.  Since $R$ is self-conjugate, if $\l$ is a weight, then so is $-\l$.  When restricted to CB flat directions \eqref{supotl} becomes
\begin{align}\label{}
\int d^2\th \sum_{\l\in R} \l(\Phi)\,  Q_{-\l,\a} J_{\a\b} Q_{\l,\b},
\end{align}
implying that for generic $\Phi$ (point on the CB) there is no potential for those $Q_\l$ components with $\l=0$.  Thus there is an ECB iff the hypermultiplet representation has a zero weight, and the hyperk\"ahler ECB fiber is given by the vevs of all the half-hypermultiplet components with $\l=0$.  There are no further gauge identifications on these components: such identifications come from the action of the Weyl group on the weights, but $\l=0$ is trivially fixed by the whole Weyl group.  Thus the fiber is $\simeq \H^h$ where $2h$ is the number of $Q_{\l=0,\a}$ half-hypermultiplet components.  $2h$ is even since a necessary condition for an irreducible representation to have a zero weight is that it be an orthogonal (a.k.a.\ real) representation,\footnote{E.g., proposition G in section 3.11 of \cite{s90} implies that if an irreducible representation has a zero weight then it is orthogonal.}  and for orthogonal representations $J_{\a\b}$ is antisymmetric and non-degenerate, and thus these representations have even multiplicity.  This is reflected in the $2n_k$ multiplicity of orthogonal representations in \eqref{Rdecomp} and the symplectic flavor symmetry factors in \eqref{ffdecomp}.  

Though zero weights only occur in orthogonal irreducible representations, not all orthogonal irreps have zero weights.  For example, the $N$-fold antisymmetric tensor product of $\SU(2N)$ fundamental representations, the vector representation of $\SO(2N)$, and the spinor representations of $\SO(4N)$ are all examples of orthogonal irreps with no zero weights.  Indeed, the weight lattice of a semi-simple Lie algebra, $\gf$, can be decomposed into disjoint affine sublattices under the action of the center of its associated unique simply-connected compact Lie group.  The irreps of $\gf$ are then organized into center conjugacy classes according to which sublattice their weights belong.  The only sublattice containing a zero weight is the root lattice, which corresponds to the center conjugacy class of the adjoint representation of $\gf$.  This conjugacy class comprises  the orthogonal representations which have zero weights.  Examples are the adjoint irrep of any simple $\gf$, the vector  irrep of $\SO(2N+1)$, the traceless-symmetric irrep of $\SO(N)$, the traceless-antisymmetric irrep of $\Sp(2N)$, and all irreps of $G_2$, $F_4$ and $E_8$.

Since ECB hypermultiplets are in orthogonal gauge representations they will be acted on by the $\Sp(2n_{k'})$ factors of $\ff$ in \eqref{ffdecomp} for those $\{k'\}\subset\{k\}$ corresponding to irreducible orthogonal representations with zero weights.  Say the $R^O_{k'}$ orthogonal irreducible representation has $q_{k'}$ zero weights.  Then the $2 q_{k'} n_{k'}$ hypermultiplet complex scalars transform under the $\Sp(2n_{k'})$ factor of $\ff$ as $q_{k'}$ copies of its $2 n_{k'}$-dimensional (``fundamental") representation (which is itself a symplectic representation).  Thus this $\Sp(2n_{k'})$ factor of $\ff$ acts on the $k'$th factor of $\H^h \sim \prod_{k'} \H^{q_{k'}n_{k'}}$ as $USp(2n_{k'})\otimes I_{q_k'}$ matrices multiplying vectors in $\C^{2n_{k'}}\otimes \C^{q_{k'}}$.  So, we have learned that in lagrangian theories ECB's can only occur in theories with symplectic flavor factors, and the ECB fiber always transforms as a direct sum of fundamental representations of (some subset of) these symplectic factors.

Now consider how the ECB fiber degenerates as we approach a singularity in the CB of a gauge SCFT.  Our earlier general discussion of the ECB fiber over a singularity showed that it can degenerate into a cone $\H^h/\sim_\s$ where $\s$ is a triholomorphic isometry of $\H^h$ which fixes the origin and commutes with the flavor action.   Consider just the component, $\H^{n_{k'} q_{k'}} \subset \H^h$, which transforms as $q_{k'}$ copies of the fundamental $2n_{k'}$-dimensional representation of the $USp(2n_{k'})$ factor of the flavor group.  Thus the most general twist upon traversing a non-trivial cycle in the CB for this factor is given by some $\s \in O(q_{k'},\C)$, the complex orthogonal group.  (In the case $q_{k'}=1$, this coincides with the $\Z_2$ center of $USp(2n_{k'})$.)  

In fact, for lagrangian SCFTs, the twist $\s$ is always in the $\Z_2$  center of the appropriate $USp(2n_{k'})$ flavor factor.  We can see this as follows.  Recall that the CB has singularities along complex co-dimension 1 subvarieties where a charged state becomes massless.  Massless hypermultiplets correspond classically to those points of the CB where the vector multiplet scalar vev $\Phi$ --- which is in a complexified Cartan subalgebra of the gauge algebra --- is annihilated by some hypermultiplet weight $\l$:  $\l(\Phi)=0$.  This classical picture of the hypermultiplet singularities is accurate in the weak coupling limit of a lagrangian SCFT, and persists to strong coupling by analytic continuation in the coupling constant.  Classically there are also singularities when $\a(\Phi)=0$ for some root $\a$, corresponding to subvarieties along which a charged vector multiplet becomes massless and an $\SU(2)$ gauge factor is restored.  These vector multiplet singularities generically flow to strong coupling where quantum effects replace them by a pair of dyonic hypermultiplet singularities \cite{sw1}.

When there is an ECB, however, we have seen that there are hypermultiplets in orthogonal gauge representations, $O_{k'}$, in the same center conjugacy class as the adjoint representation.  In this case the weights of $O_{k'}$ hypermultiplets are in the root lattice, and for SCFTs they are proportional to the roots themselves.\footnote{In general, elements of the root lattice need not be proportional to roots.  However, all the representations with weights in the root lattice which can occur in a SCFT (i.e., do not give too large a contribution to the beta function) have weights which are a subset of the roots, as can be checked by inspection, e.g., using the classification of gauge SCFTs given in \cite{Bhardwaj:2013qia}.  This discussion can be extended to IR-free gauge theories, in which case the $O_{k'}$ irreps may have weights in the root lattice which are not proportional to any root.  The singularities where components of the $O_{k'}$ hypermultiplet with such weights become massless will then not have an enhanced $\SU(2)$ IR gauge symmetry, and so no $\Z_2$ identification on the ECB fiber.}  Thus the singularities $\l(\Phi)=0$ where a component of an $O_{k'}$ hypermultiplet becomes massless coincide with the singularities $\a(\Phi)=0$ where classically an IR $\SU(2)$ gauge symmetry is restored since $\l=\a$.  This IR $\SU(2)$ theory therefore has the representation content of an $\cN=2$ $\SU(2)$ gauge theory with a positive number (the multiplicity $q_{k'}n_{k'}$ of the weight $\l$) of massless adjoint hypermultiplets.  It will thus be either a scale invariant or IR-free $\SU(2)$ theory, and so, in particular, there will be no quantum corrections to the classical description of its singularity. 

Because the Weyl group $\cW$ acts nontrivially on the Cartan subalgebra, $\Phi$ is not a gauge-invariant coordinate on the CB.  The Weyl group is generated by reflections through hyperplanes annihilated by the gauge algebra roots.  Thus a loop in the CB linking the singular subvariety corresponding to $\a(\Phi)=0$ is lifted to an open path in the complexified Cartan subalgebra connecting a point $\Phi_*$ to its image under the Weyl reflection through the $\a(\Phi)=0$ hyperplane.  This Weyl reflection is the element of the enhanced $\SU(2)$ gauge group which acts as on the neutral massless hypermultiplets there as $\s : z_k \mapsto -z_k$ for each $z_k\in\C^{n_k}$, which is the action of the center of $USp(2n_{k'})$.

\providecommand{\href}[2]{#2}\begingroup\raggedright\endgroup

\end{document}